\documentclass[12pt]{article}

\usepackage[reqno]{amsmath}
\usepackage{mathrsfs}
\usepackage{amssymb}

\usepackage{rotating} %for landscape format

\usepackage{array}
\usepackage{float}
\usepackage{color}
\usepackage{graphicx}% Include figure files
\usepackage{a4}
\usepackage{a4wide}
\usepackage{wasysym}
\def\gs{\mathrel{
   \rlap{\raise 0.511ex \hbox{$>$}}{\lower 0.511ex \hbox{$\sim$}}}}
\def\ls{\mathrel{
   \rlap{\raise 0.511ex \hbox{$<$}}{\lower 0.511ex \hbox{$\sim$}}}}

\newcommand{\be}{\begin{eqnarray}}
\newcommand{\ee}{\end{eqnarray}}
\newcommand{\beq}{\begin{equation}}
\newcommand{\eeq}{\end{equation}}
\newcommand{\bena}{\begin{eqnarray}}
\newcommand{\eena}{\end{eqnarray}}

\newcommand{\Rep}[1]{\underline{\mbox{\textbf{#1}}}}

\begin{document}

%For feynmf-Package:
\setlength{\unitlength}{1mm}

\begin{titlepage}
\title{\vspace*{-2.0cm}
%\hfill {\small hep--ph/xxxxxx}\\[20mm]
\bf\Large
Deriving Models for keV sterile Neutrino Dark Matter with the Froggatt-Nielsen mechanism
\\[5mm]\ }

\author{
Alexander Merle$^a$\thanks{email: \tt amerle@kth.se}~~~and~~
Viviana Niro$^{bc}$\thanks{email: \tt niro@to.infn.it}
\\ \\
$^a${\normalsize \it Department of Theoretical Physics, School of Engineering Sciences,}\\
{\normalsize \it Royal Institute of Technology (KTH), AlbaNova University Center,}\\
{\normalsize \it Roslagstullsbacken 21, 106 91 Stockholm, Sweden}\\
\\
$^b${\normalsize \it Dipartimento di Fisica Teorica, Universit\`a di Torino}\\
{\normalsize \it and INFN, Sez. di Torino, via P.\ Giuria 1, I-10125 Torino, Italy}\\
\\
$^c${\normalsize \it Max-Planck-Institut f\"ur Kernphysik,}\\
{\normalsize \it Postfach 10 39 80, 69029 Heidelberg, Germany}
}
\date{\today}
\maketitle
\thispagestyle{empty}

\begin{abstract}
\noindent
Sterile neutrinos with a mass around the keV scale are an attractive particle physics candidate for Warm Dark Matter. 
Although many frameworks have been presented in which these neutrinos can fulfill all phenomenological constraints, 
there are hardly any models known that can explain such a peculiar mass pattern, one sterile neutrino at the keV scale 
and the other two considerably heavier, while at the same time being compatible with low-energy neutrino data. 
In this paper, we present models based on the Froggatt-Nielsen mechanism, which can give such an explanation. 
We explain how to assign Froggatt-Nielsen charges in a successful way, and we give a detailed discussion of all 
conditions to be fulfilled. It turns out that the typical arbitrariness of the charge assignments is greatly reduced 
when trying to carefully account for all constraints. We furthermore present analytical calculations of a few simplified 
models, while quasi-perfect models are found numerically. 
\end{abstract}

\end{titlepage}

%%%%%%%%%%%%%%%%%%%%%%%%%%%%%%%%%%%%%%%%%%%%%%%%%%%%%%%%%%%%%%%%%%%%%%
\section{\label{sec:intro} Introduction}
%%%%%%%%%%%%%%%%%%%%%%%%%%%%%%%%%%%%%%%%%%%%%%%%%%%%%%%%%%%%%%%%%%%%%%

One of the most intriguing problems in today's particle physics and astrophysics is the identity of the mysterious 
Dark Matter~\cite{Zwicky:1933gu} observed in the Universe~\cite{Komatsu:2010fb}. It is known since a few years
 that computer simulations of small scale structure formation are in favour of so-called ``warm'' Dark Matter (WDM), 
with a mass of roughly 1--2~keV~\cite{Bode:2000gq,deVega:2010yk}, which is an intermediate case between the standard paradigm 
of ``cold'' (non-relativistic) and ``hot'' (relativistic) Dark Matter, the latter being strongly disfavored by structure 
formation. Furthermore, there are model-independent data analyses which also seem to point to the keV scale~\cite{deVega:2009ku}.

A very interesting framework providing a WDM candidate motivated by particle physics is the so-called 
$\nu$MSM~\cite{Asaka:2005an}, which extends the Standard Model (SM) of particle physics by three right-handed (sterile) 
neutrinos, one of which has a mass at the keV scale, whereas the other two are considerably heavier. However, it turns out 
that, in the standard thermal production, the Dark Matter would be overproduced, which makes it necessary to rely on 
non-thermal production instead~\cite{Shaposhnikov:2006xi,Bezrukov:2008ut,Bezrukov:2009yw}. An alternative is provided by embedding the 
keV sterile neutrino in a gauge extension of the SM and correcting the abundance by allowing for sufficient entropy 
production in the decay of the two heavier sterile neutrinos, which has been exemplified in a Left-Right symmetric 
framework~\cite{Bezrukov:2009th}. Note, however, that this last proposal required a seesaw type~II situation in order 
not to be in conflict with the experimental and observational constraints. In addition, keV sterile neutrinos can also 
appear in the frameworks of the scotogenic/inert Higgs doublet model~\cite{Sierra:2008wj,Gelmini:2009xd} or of composite 
neutrinos~\cite{Grossman:2010iq}.

The common feature of all these proposals is, however, that they do not yield an {\em explanation} of the required 
mass pattern of sterile neutrinos: They can be seen as very useful frameworks providing all tools to make concrete 
predictions, and the corresponding parameters can assume values leading to full agreement with data. Nevertheless, 
in the view of model building, they only assume the correct mass pattern to be present.

\begin{figure}[t]
\centering
\begin{tabular}{cc}
\includegraphics[width=6.0cm]{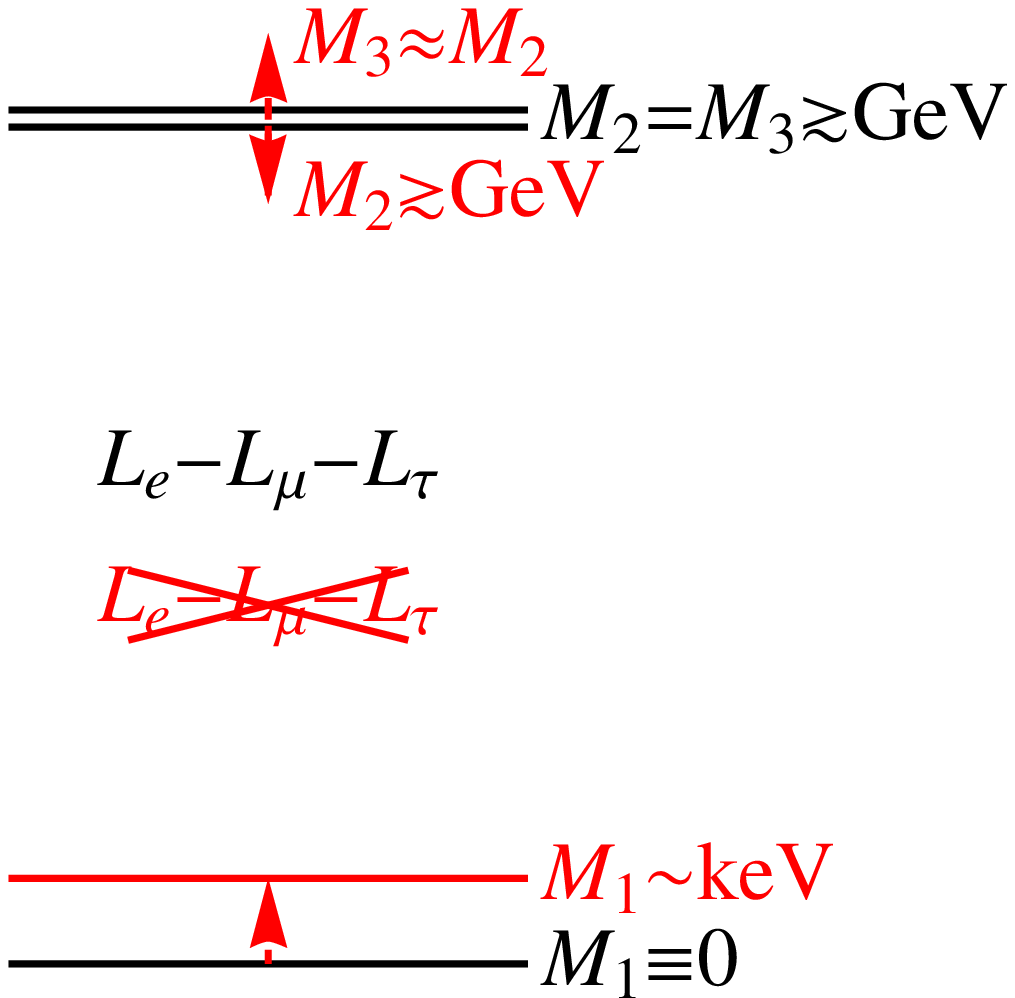}& 
\includegraphics[width=8.5cm]{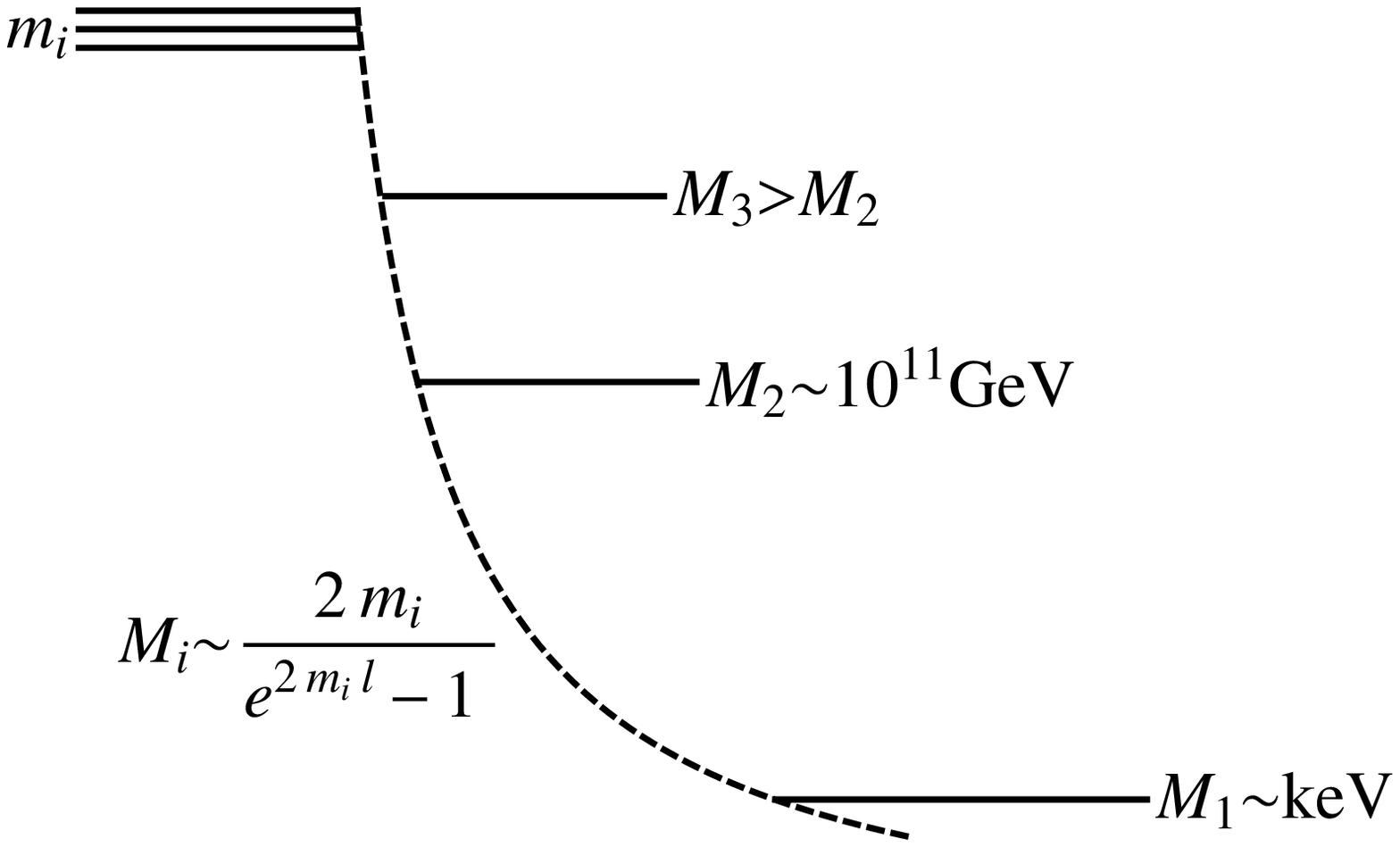}
\end{tabular}
\caption{\label{fig:schemes} The mass shifting schemes of the two classes of models that had already existed: The left panel 
displays the shifting due to soft breaking of a global $L_e-L_\mu-L_\tau$ flavour symmetry, while the right one shows the 
exponential suppression in a Randall-Sundrum framework.}
\end{figure}

Up to now, to our knowledge, only two classes models have existed which could yield an explanation of the sterile neutrino 
mass pattern. The key point is to achieve a strong mass splitting in the right-handed neutrino sector, as only the 
lightest sterile particle should have a mass in the keV range, while the other two must be considerably 
heavier~\cite{Bezrukov:2009th}. This leads to different schemes for shifting a certain initial mass spectrum, two of which are 
depicted in Fig.~\ref{fig:schemes}. The first class of models is based on a flavour symmetry which forces one sterile neutrino 
to be strictly massless. This idea has first been discussed in Ref.~\cite{Shaposhnikov:2006nn} for a specifically defined lepton 
number symmetry, and has recently been investigated in the context of a $L_e-L_\mu-L_\tau$ flavour symmetry~\cite{Lindner:2010wr}. 
Both symmetries can explain such a peculiar pattern of right-handed neutrino masses easily: The trick is to make use of the 
different scales of symmetry breaking and symmetry preserving terms to generate the necessary hierarchy. Soft breaking of the 
symmetry will lift up the mass of one sterile neutrino, which would have been strictly massless in the limit of restored 
symmetry, to the keV scale, see left panel of Fig.~\ref{fig:schemes} for the $L_e-L_\mu-L_\tau$ case. The breaking will 
furthermore break up the predicted exact mass degeneracy in the light neutrino sector, which would otherwise contradict the data. 
The second real model~\cite{Kusenko:2010ik} exploited the exponential factor in a Randall-Sundrum~\cite{Randall:1999ee} like framework 
to obtain a large mass splitting from very moderately tuned parameters, cf.\ right panel of Fig.~\ref{fig:schemes}. This model 
has the advantages of naturally explaining the existence of one sterile neutrino at the keV scale, while the other two could 
have masses of around $10^{11}$~GeV or even heavier, and of at the same time insuring that the seesaw 
mechanism~\cite{Minkowski:1977sc,Yanagida:1979as,GellMann:1980vs,Glashow:1979nm,Mohapatra:1979ia} works, in spite of the 
existence of a relatively light sterile neutrino. On the other hand, this model involves the assumption of a suitable UV-brane, 
which is intrinsically hard to probe.

In this paper, we want to present a third alternative to explain the existence of a keV scale sterile neutrino, by making 
use of the Froggatt-Nielsen (FN) mechanism~\cite{Froggatt:1978nt}. This mechanism is well-known to be capable of explaining 
very strong mass hierarchies, e.g.\ in the quark sector~\cite{Buchmuller:1998zf,Altarelli:1998ns,Kamikado:2008jx}, which is 
just what is needed for having one sterile neutrino at the keV scale whereas the other two are heavier by at least six orders 
of magnitude. Our goal is to find the {\it minimal} assignments needed to explain the pattern in the sterile neutrino sector, 
while being in full agreement with the rest of the lepton data. The key point of the FN mechanism is that at most 
one fermion generation (namely the third one, if at all) obtains a mass by the standard Higgs 
mechanism, or by the existence of a direct mass term for gauge singlets. The other generations, however, will only receive 
higher-order contributions from multiple seesaw-like diagrams, which leads to the typical cascade or waterfall 
structures~\cite{Haba:2008dp}. This is enforced by having an additional Abelian $U(1)_{\rm FN}$ flavour symmetry under which 
the first, second, and third generations have charges with decreasing absolute values. Furthermore, such a framework requires 
a heavy sector of new fermions that is usually not specified in details, as well as so-called flavon fields, which are SM 
singlets but charged under the $U(1)_{\rm FN}$. These fields obtain vacuum expectation values (VEVs) in order to break the 
symmetry in a phenomenologically suitable way~\cite{Chen:2003zv}. Integrating out the heavy fields finally leads to suppression 
factors by some power of a small parameter $\lambda$, which is usually taken to be of the order of the Cabibbo 
angle~\cite{Datta:2005ci}.

Note that, shortly before this work was completed, Ref.~\cite{Barry:2011wb} appeared, where a Froggatt-Nielsen mechanism, 
together with a $Z_3$ symmetry, is used in the context of an $A_4$ flavour symmetry to generate a hierarchy for the charged 
lepton masses and to regulate the mass of a sterile neutrino to be around the eV-scale. The authors also comment on the 
possibility of exploiting this model for an explanation of a sterile neutrino with a keV mass, however, without investigating 
this route explicitly. In their case, however, the Froggatt-Nielsen mechanism is merely a small addendum to the existing 
$A_4$ model, whereas we consider the mechanism as starting point to develop fully working models.

The structure of this paper is the following: In Sec.~\ref{sec:FN} the main features of the Froggatt-Nielsen mechanism 
are summarized The possible choices for the FN charges of the right-handed neutrinos are analyzed in Sec.~\ref{sec:MR}. 
After that, in Sec.~\ref{sec:problems}, we give a detailed discussion about all requirements to be fulfilled, where we will 
also explain why certain frameworks are incompatible with the FN mechanism when aiming at a description of keV mass sterile 
neutrinos. In addition, we will also discuss why certain constraints that might be problematic for a higher right-handed 
neutrino mass scale are practically of no relevance to our case. Finally, in Sec.~\ref{sec:SU(5)}, we present analytical 
calculations of some promising scenarios, as well as fully working numerical models, before concluding in Sec.~\ref{sec:conc}. 
Explicit expressions for the diagonalization matrices used can be found in Appendix~A.

%%%%%%%%%%%%%%%%%%%%%%%%%%%%%%%%%%%%%%%%%%%%%%%%%%%%%%%%%%%%%%%%%%%%%%
\section{\label{sec:FN} The Froggatt-Nielsen mechanism}
%%%%%%%%%%%%%%%%%%%%%%%%%%%%%%%%%%%%%%%%%%%%%%%%%%%%%%%%%%%%%%%%%%%%%%

The Froggatt-Nielsen (FN) mechanism~\cite{Froggatt:1978nt} is maybe one of the best possibilities 
to explain strong hierarchies between fermion masses. Furthermore, there exist successful explicit 
applications of this mechanism to the neutrino sector, see, e.g., Refs.~\cite{Kamikado:2008jx,Kane:2005va}, 
and Refs.~\cite{Plentinger:2006nb,Plentinger:2007px,Plentinger:2008up,Niehage:2008sg} for a systematic parameter space scan.

We would like to use the FN mechanism to explain a hierarchical 
spectrum (HS) in the sterile neutrino sector of the following type:
$M_1 \simeq \mathcal{O}({\rm keV})$ and $M_2, M_3 \gtrsim \mathcal{O}({\rm GeV})$. 
Denoting the FN flavon field by $\Theta$, and the cut-off scale at which the heavy 
sector of the theory is integrated out by $\Lambda$, we can write the Lagrangian 
as: 

\bena
\mathcal{L}_{\rm leptons}&=&
-Y^{i j}_e\,\overline{e_{i R}}\,H\,L_{j L}\,
\left( \frac{\Theta}{\Lambda} \right)^{k_i+f_j} + h.c.
-Y^{i j}_D\,\overline{N_{i R}}\,\tilde{H}\,L_{j L}\,
\left( \frac{\Theta}{\Lambda} \right)^{g_i+f_j} + h.c. \label{eq:lepton_masses}\\
&&-\frac{1}{2}\,\overline{N_{i R}}\,\tilde{M}^{i j}_R\,(N_{j R})^C
\left( \frac{\Theta}{\Lambda} \right)^{g_i+g_j} + h.c.
-\frac{1}{2}\,Y^{i j}_L\,\overline{(L_{i L})^C}\,(i \sigma_2 \Delta)\,L_{j L}
\left( \frac{\Theta}{\Lambda} \right)^{f_i+f_j} + h.c.\nonumber\,,
\eena 
where $H$ is the Standard Model Higgs and $\tilde{H}=i \sigma_2 H^*$ is its charge 
conjugate (see Tab.~\ref{tab:table_quantumnumbers} for the definition 
of all the other fields). 
The FN flavon field can acquire a VEV $\langle \Theta \rangle$, with 
$\lambda =\frac{\langle \Theta \rangle}{\Lambda}$ being a small 
quantity of the order of the Cabibbo angle: $\lambda \simeq 0.22$~\cite{Datta:2005ci}. 

Note that, in principle, 
FN charges could be positive or negative, which would correspond to integrating out heavy particles or 
anti-particles, respectively. This would again lead to suppression factors due to having a 
higher mass scale involved. Accordingly, the decisive quantity for the suppression is actually the 
absolute value of the sum of FN charges. 
After the flavon field acquires a VEV and after electroweak symmetry breaking, the charged leptons 
and the Dirac neutrino mass matrices are given by
\begin{table}[t]
 \centering
 \begin{tabular}{|c||c|c|c|c|c|c|}\hline
 Field & $L_{i L}$ & 
 $\overline{e_{i R}}$ & 
 $\overline{N_{i R}}$ & 
 $H$ & $\Delta$ & $\Theta$ \\\hline\hline
 $SU(2)_L$ & 
 \Rep{2} & \Rep{1} & \Rep{1} & \Rep{2} & \Rep{3} & \Rep{1}\\ \hline
 $U(1)_{\rm FN}$ & $f_i$ & 
 $k_i$ &
 $g_i$ & 0 & 0 & $-1$\\\hline
 \end{tabular}
 \caption{\label{tab:table_quantumnumbers} $SU(2)$ and family charges, with $i=1,2,3$.}
\end{table}

\begin{equation}
M_e=v
\begin{pmatrix}
Y^{11}_e\,\lambda^{|k_1+f_1|} & Y^{12}_e\,\lambda^{|k_1+f_2|} & Y^{13}_e\,\lambda^{|k_1+f_3|}\\
Y^{21}_e\,\lambda^{|k_2+f_1|} & Y^{22}_e\,\lambda^{|k_2+f_2|} & Y^{23}_e\,\lambda^{|k_2+f_3|}\\
Y^{31}_e\,\lambda^{|k_3+f_1|} & Y^{32}_e\,\lambda^{|k_3+f_2|} & Y^{33}_e\,\lambda^{|k_3+f_3|}
\end{pmatrix}\,,
\label{eq:Ml}
\end{equation}
\begin{equation}
m_D=v
\begin{pmatrix}
Y^{11}_D\,\lambda^{|g_1+f_1|} & Y^{12}_D\,\lambda^{|g_1+f_2|} & Y^{13}_D\,\lambda^{|g_1+f_3|}\\
Y^{21}_D\,\lambda^{|g_2+f_1|} & Y^{22}_D\,\lambda^{|g_2+f_2|} & Y^{23}_D\,\lambda^{|g_2+f_3|}\\
Y^{31}_D\,\lambda^{|g_3+f_1|} & Y^{32}_D\,\lambda^{|g_3+f_2|} & Y^{33}_D\,\lambda^{|g_3+f_3|}
\end{pmatrix}\,,
\label{eq:mD}
\end{equation}
where $v$ is the VEV of the Higgs doublet and $f_i$, $k_i$, and $g_i$ are the FN charges 
of the left-handed lepton doublets, the right-handed charged, and the right-handed neutrinos, respectively, see 
Tab.~\ref{tab:table_quantumnumbers}. 

Now let us have a look at the right-handed neutrino sector. 
Since the corresponding mass matrix $M_R$ has to be symmetric, it will have the following structure:
\begin{equation}
M_R=
\begin{pmatrix}
\tilde{M}^{11}_R\,\lambda^{|2 g_1|}\hfill \hfill & \tilde{M}^{12}_R\,\lambda^{|g_1+g_2|} & \tilde{M}^{13}_R\,
\lambda^{|g_1+g_3|}\\
\bullet & \tilde{M}^{22}_R\,\lambda^{|2 g_2|}\hfill\hfill & \tilde{M}^{23}_R\,\lambda^{|g_2+g_3|}\\
\bullet & \bullet & \tilde{M}^{33}_R\,\lambda^{|2 g_3|}\,\hfill \hfill 
\end{pmatrix}\,.
 \label{eq:MR_struc}
\end{equation}
Applying the type~I seesaw formula then leads to the following structure:
\begin{equation}
m^I_\nu=
-m^T_D\,M^{-1}_R\,m_D=
\begin{pmatrix}
a_1\,\lambda^{|2 f_1|} & b_1\,\lambda^{|f_1+f_2|} & c_1\,\lambda^{|f_1+f_3|}\\
\bullet & d_1\,\lambda^{|2 f_2|} & e_1\,\lambda^{|f_2+f_3|}\\
\bullet & \bullet & f_1\,\lambda^{|2 f_3|}
\end{pmatrix}\,,
\label{eq:typeI}
\end{equation}
where the parameters $a_1,b_1,c_1,d_1,e_1,f_1$ depend on the parameters present 
in the mass matrices $m_D$ and $M_R$. In the case of a type~II seesaw scenario, one needs the left-handed 
Majorana neutrino mass in addition,
\begin{equation}
m_L=v_\Delta
\begin{pmatrix}
Y^{11}_L\,\lambda^{|2 f_1|} \hfill \hfill & Y^{12}_L\,\lambda^{|f_1+f_2|} & Y^{13}_L\,\lambda^{|f_1+f_3|}\\
\bullet & Y^{22}_L\,\lambda^{|2 f_2|} \hfill \hfill & Y^{23}_L\,\lambda^{|f_2+f_3|}\\
\bullet & \bullet & Y^{33}_L\,\lambda^{|2 f_3|}\,\hfill \hfill
\end{pmatrix}\,,
\label{eq:mL}
\end{equation}
with $v_\Delta$ being the VEV of the triplet $\Delta$. The type~II seesaw neutrino mass matrix 
will be given by 
\begin{equation}
m^{II}_\nu=
m_L-m^T_D\,M^{-1}_R\,m_D=
m_L+m^I_\nu
=
\begin{pmatrix}
a_2\,\lambda^{|2 f_1|} & b_2\,\lambda^{|f_1+f_2|} & c_2\,\lambda^{|f_1+f_3|}\\
\bullet & d_2\,\lambda^{|2 f_2|} & e_2\,\lambda^{|f_2+f_3|}\\
\bullet & \bullet & f_2\,\lambda^{|2 f_3|}
\end{pmatrix}\,,
\label{eq:typeII}
\end{equation}
where the parameters $a_2,b_2,c_2,d_2,e_2,f_2$ depend on the parameters present 
in the mass matrices $m_{L}$, $m_D$, and $M_R$.

Note that Eqs.~\eqref{eq:typeI} and~\eqref{eq:typeII} proof that the seesaw mechanism is, in this framework, 
not spoiled by the presence of a keV neutrino: Any global $U(1)$ charge would cancel out in the seesaw formula, 
and so do the $U(1)_{\rm FN}$ charges of the right-handed fermions. These charges are, however, the only 
connection to the keV scale, since they lower the natural scale $M_0$ of the right-handed neutrino mass by 
introducing certain powers of $\lambda$. Accordingly, in the seesaw formula, one is always guaranteed to divide by a 
relatively large value $M_0$, thereby saving the seesaw mechanism. In fact, since the FN charges of the left-handed 
lepton doublets are still present in the seesaw formula, the suppression mechanism for the light neutrino masses might 
even be amplified.

Furthermore one can see from Eqs.~\eqref{eq:typeI} and~\eqref{eq:typeII} that the structure of the neutrino mass 
matrices is precisely the same for type~I and type~II seesaw scenarios, which is another consequence of the right-handed 
neutrino charges canceling out. For this reason, we will not consider type~II seesaw scenarios any further in this paper, 
since the results for such scenarios can be trivially recovered from the seesaw type~I results, by the simple transformations 
$a_1\to a_2$, $b_1\to b_2$, and so on. We will, however, mention differences or additional features that would appear in 
the type~II seesaw case, where appropriate.

Before starting to apply the FN mechanism to explain keV sterile neutrinos, we also want to stress a potential problem of this method: The Froggatt-Nielsen mechanism intrinsically involves a high energy sector that is not specified further. This is a problem not only of this particular mechanism, but of practically all flavour symmetries that must be broken in a phenomenologically acceptable way. Such a breaking always involves a scalar (flavon) sector, which is assumed to have suitable properties. In addition, the FN mechanism also involves heavy fermions, which are integrated out to lead to the suppression factors in Eq.~\eqref{eq:lepton_masses}. Such high energy sectors, although of no practical relevance for low energy experiments, could potentially become important for very high energies, and hence in particular in the early Universe. This problem, however, is strongly model-dependent, and far beyond the scope of a conceptual paper like the one presented here.

We nevertheless want to point out that the additional interactions generated by this high energy sector could, e.g., cause the keV sterile neutrinos to be in thermal equilibrium at early times. Although certain production mechanisms of keV sterile neutrino DM require the sterile neutrinos to never enter thermal equilibrium (see, e.g., Refs.~\cite{Asaka:2005an,Shaposhnikov:2006xi,Bezrukov:2008ut,Bezrukov:2009yw}), there is also the alternative possibility to first produce them thermally and then dilute their abundance by sufficient entropy production~\cite{Bezrukov:2009th}. The lesson to learn is that one should be careful when applying the FN mechanism to certain scenarios: We will discuss many accompanying problems from the particle physics side later on in Sec.~\ref{sec:problems}. However, one has to keep in mind that also certain astrophysical scenarios could lead to further restrictions or incompatibilities.

%%%%%%%%%%%%%%%%%%%%%%%%%%%%%%%%%%%%%%%%%%%%%%%%%%%%%%%%%%%%%%%%%%%%%%
\section{\label{sec:MR} FN charges for the right-handed neutrinos}
%%%%%%%%%%%%%%%%%%%%%%%%%%%%%%%%%%%%%%%%%%%%%%%%%%%%%%%%%%%%%%%%%%%%%%

The first question we want to answer is how to assign FN charges to the right-handed neutrino 
fields. The most important constraint to keep in mind is the
strong hierarchy in the right-handed (sterile) neutrino sector: In a framework as in
Ref.~\cite{Bezrukov:2009th}, where the lightest heavy neutrino is supposed to have a
mass $M_1$ of a few keV, the second to lightest (heavy) neutrino must at least have a
mass $M_2$ of order GeV, due to the requirement of sufficient entropy production.
This condition does, however, not constrain the largest mass $M_3$, which can hence
be similar to or even much larger than $M_2$. In any case, we require a hierarchy of at least six
orders of magnitude between $M_1$ and $M_2$. 

If we suppose that all the coefficients in the matrix $M_R$ are of the same order, then 
the explanation of the HS in the sterile neutrino sector should come 
from the FN charges. 
Considering the FN parameter $\lambda$ to be of the size of the Cabibbo angle~\cite{Datta:2005ci} 
($\lambda \simeq 0.22$) and $g_1 \geq g_2 \geq g_3$, the minimal conditions to be fulfilled are:
\be
\left\{
\begin{array}{rclcrcl}
 g_1 &\geq& g_1|_{\rm min} &\quad {\rm{ with  }} \qquad& g_1|_{\rm min}&=&g_2+3\,,\nonumber\\
 g_2 &\geq& g_2|_{\rm min} &\quad {\rm{ with  }} \qquad& g_2|_{\rm min}&=&g_3\,.
 \label{eq:FN-condition}
\end{array}
\right.
\ee
In the minimal case, 
$g_2 = g_3 = g_1-3$, we would obtain a mass spectrum of the following type: 
$M_1 \simeq \mathcal{O}({\rm keV})$ and $M_2, M_3 \simeq \mathcal{O}({\rm GeV})$. 
For $g_2 = g_1-3$ and $g_3 < g_2$, the mass spectrum would instead be given 
by: 
$M_1 \simeq \mathcal{O}({\rm keV})$, $M_2\simeq \mathcal{O}({\rm GeV})$, and 
$M_3 > \mathcal{O}({\rm GeV})$. 
Finally, if $g_2 < g_1-3$ and $g_3 < g_2$, both $M_2$ and $M_3$ have 
a mass greater than $\mathcal{O}({\rm GeV})$.  

Since we want to stick to minimal choices of FN charge assignments, a reasonable condition 
would be to set $g_3=0$ and $g_1=g_2+3$. Accordingly, we choose two minimal scenarios that 
we will investigate further:
\begin{itemize}
 \item Scenario~A: $(g_1,g_2,g_3)=(3,0,0)$,
 \item Scenario~B: $(g_1,g_2,g_3)=(4,1,0)$.
\end{itemize}
These scenarios lead to mass eigenvalues $M_{1,2,3}$ which obey the hierarchies $M_1\approx 10^{-6} M_{2,3}$ 
and $M_1\approx 10^{-6} M_2 \approx 10^{-8} M_3$, respectively. 
In Tabs.~\ref{tab:table_300} and~\ref{tab:table_410}, we show some examples of 
allowed textures that lead to the desired HS, considering two, three, or 
four independent parameters in the mass matrix $M_R$.

\begin{table}[t]
 \centering
 \begin{tabular}{|c||c|}\hline
$M_R$ & Eigenvalues \\\hline\hline
$\begin{pmatrix}
A\,\lambda^{6} & A\,\lambda^{3} & A\,\lambda^{3}\\
A\,\lambda^{3} & A & B\\
A\,\lambda^{3} & B & A
\end{pmatrix}$ & 
$\begin{matrix}
M_1 = & \mathcal{O}(\lambda^6)\simeq \mathcal{O}({\rm keV})\\
M_2 = & A-B \simeq \mathcal{O}({\rm GeV})\\
M_3 = & A+B \simeq \mathcal{O}({\rm GeV})
\end{matrix}$  \\ \hline
$\begin{pmatrix}
A\,\lambda^{6} & A\,\lambda^{3} & A\,\lambda^{3}\\
A\,\lambda^{3} & B & C\\
A\,\lambda^{3} & C & B
\end{pmatrix}$ & 
$\begin{matrix}
M_1 = & \mathcal{O}(\lambda^6)\simeq \mathcal{O}({\rm keV})\\
M_2 = & B-C \simeq \mathcal{O}({\rm GeV})\\
M_3 = & B+C \simeq \mathcal{O}({\rm GeV})
\end{matrix}$ \\ \hline
$\begin{pmatrix}
A\,\lambda^{6} & B\,\lambda^{3} & B\,\lambda^{3}\\
B\,\lambda^{3} & C & D\\
B\,\lambda^{3} & D & C
\end{pmatrix}$  & 
$\begin{matrix}
M_1 = & \mathcal{O}(\lambda^6)\simeq \mathcal{O}({\rm keV})\\
M_2 = & C-D \simeq \mathcal{O}({\rm GeV})\\
M_3 = & C+D \simeq \mathcal{O}({\rm GeV})
\end{matrix}$\\\hline
 \end{tabular}
 \caption{\label{tab:table_300} 
Examples of $M_R$ textures for the Scenario~A that lead to a HS. 
We have assumed all the parameters to be greater than zero.}
\end{table}

\begin{table}[t]
 \centering
 \begin{tabular}{|c||c|}\hline
$M_R$ & Eigenvalues \\\hline\hline
$\begin{pmatrix}
A\,\lambda^{8} & A\,\lambda^{5} & A\,\lambda^{4}\\
A\,\lambda^{5} & A \lambda^2 & B \lambda\\
A\,\lambda^{4} & B \lambda & A
\end{pmatrix}$ & 
$\begin{matrix}
M_1 = & \mathcal{O}(\lambda^8)\simeq \mathcal{O}({\rm keV})\\
M_2 = & \mathcal{O}(\lambda^2) \simeq \mathcal{O}({\rm GeV})\\
M_3 = & A \simeq \mathcal{O}({\rm 100\,GeV})
\end{matrix}$  \\ \hline
$\begin{pmatrix}
A\,\lambda^{8} & A\,\lambda^{5} & A\,\lambda^{4}\\
A\,\lambda^{5} & B \lambda^2 & C \lambda\\
A\,\lambda^{4} & C \lambda & B
\end{pmatrix}$ & 
$\begin{matrix}
M_1 = & \mathcal{O}(\lambda^8)\simeq \mathcal{O}({\rm keV})\\
M_2 = & \mathcal{O}(\lambda^2) \simeq \mathcal{O}({\rm GeV})\\
M_3 = & B \simeq \mathcal{O}({\rm 100\,GeV})
\end{matrix}$  \\ \hline
$\begin{pmatrix}
A\,\lambda^{8} & B\,\lambda^{5} & B\,\lambda^{4}\\
B\,\lambda^{5} & C \lambda^2 & D \lambda\\
B\,\lambda^{4} & D \lambda & C
\end{pmatrix}$  & 
$\begin{matrix}
M_1 = & \mathcal{O}(\lambda^8) \simeq \mathcal{O}({\rm keV})\\
M_2 = & \mathcal{O}(\lambda^2)\simeq \mathcal{O}({\rm GeV})\\
M_3 = & C \simeq \mathcal{O}({\rm 100\,GeV})
\end{matrix}$ \\\hline
 \end{tabular}
 \caption{\label{tab:table_410} 
Examples of $M_R$ textures for the Scenario~B that lead to a HS. 
We have assumed all the parameters to be greater than zero.}
\end{table}

As explained before, one could increase the splitting of the mass eigenvalues by assuming 
stronger hierarchies in the FN charges. Furthermore, one could 
in principle also assign a non-zero $g_3$, which would decrease the values of the masses 
compared to some characteristic scale that could, e.g., be generated by the VEV of some scalar field.

\begin{figure}[t]
\centering
\includegraphics[width=9cm]{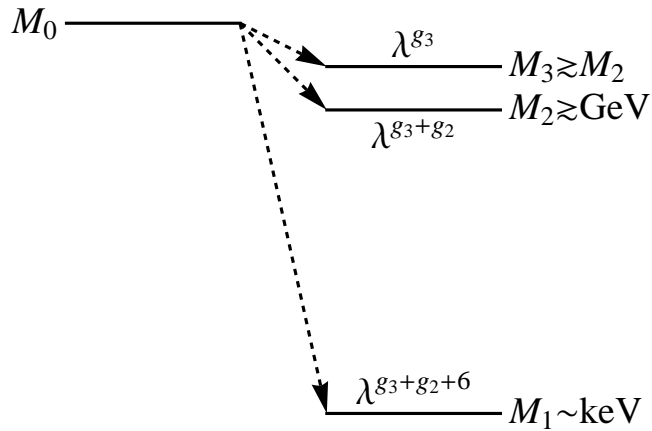}
\caption{\label{fig:FN-scheme} The mass shifting scheme of Froggatt-Nielsen models.}
\end{figure}

In Fig.~\ref{fig:FN-scheme}, we have schematically depicted the general effect of the FN mechanism: A certain mass scale 
$M_0$ is multiplied by powers of $\lambda$ that depend on the fermion generation. These factors lead to suppressions 
of the physical mass eigenvalues. In general, FN assignments are very well suited to explain strong hierarchies, which 
is why we ultimately chose to investigate this framework.

In the following, we will investigate only the two exemplifying Scenarios~A and~B, and we will show 
how to implement their assignments in a more complete model.

%%%%%%%%%%%%%%%%%%%%%%%%%%%%%%%%%%%%%%%%%%%%%%%%%%%%%%%%%%%%%%%%%%%%%%
\section{\label{sec:problems} Requirements, No-Go's, and amenities}
%%%%%%%%%%%%%%%%%%%%%%%%%%%%%%%%%%%%%%%%%%%%%%%%%%%%%%%%%%%%%%%%%%%%%%

In this section, we will explain in a careful way why several FN scenarios are not able to 
successfully produce a neutrino sector compatible with the data, or at least 
have problems with it. In fact, even though the FN framework seems to involve quite some freedom, 
the way how the charges are combined is nevertheless quite restrictive: By 
fixing only 9 charges (lepton doublets, right-handed electrons, and right-handed neutrinos), 
we predict the magnitudes of $(9+9+6)=24$ (type I seesaw) or $(9+9+6+6)=30$ (type II seesaw) 
entries in the matrices $M_e$, $m_D$, $M_R$, and $m_L$, which would otherwise be independent.

Accordingly, there can be many allowed theoretical and phenomenological scenarios that, when 
implemented within a FN framework, may lead to contradictions. In the following, we will discuss several situations 
which can lead to problems, as well as certain requirements that have to be fulfilled, thereby reducing step by step the 
arbitrariness 
involved in the FN models. Furthermore, we will point out why certain constraints that are typically 
problematic for FN inspired models do not apply in our case. As it will turn out, although FN charge assignments might 
seem relatively random at the first sight, they are in fact restrictive enough to disagree with several frameworks, and 
in particular they disagree with the Left-Right symmetric framework for keV sterile neutrino Dark Matter proposed in 
Ref.~\cite{Bezrukov:2009th}, as well as with the bimaximal mixing from the neutrino side proposed in this context in 
Ref.~\cite{Lindner:2010wr}. Turning this round, it will be easy to rule out the FN framework if evidence for one of the 
contradicting scenarios is found.

Note that one could, in principle, apply additional flavour symmetries 
to force some entries in the mass matrices to zero. This method could alter all 
conclusions in this section, but such an approach is, on the other hand, far away from being 
minimalistic. In particular, the FN mechanism might not even be necessary anymore in such 
cases if, e.g., some VEV hierarchies are imposed or if the symmetry is softly broken. Since 
we want to stick to FN situations, however, we do not consider such extended scenarios in this paper.

%%%%%%%%%%%%%%%%%%%%%%%%%%%%%%%%%%%%%%%%%%%%%%%%%%%%%%%%%%%%%%%%%%%%%%
\subsection{\label{sec:bimax} No bimaximal neutrino mixing}
%%%%%%%%%%%%%%%%%%%%%%%%%%%%%%%%%%%%%%%%%%%%%%%%%%%%%%%%%%%%%%%%%%%%%%

Let us now investigate the compatibility of certain scenarios with the FN framework. One easy scenario that leads to a 
tri-bimaximal PMNS matrix~\cite{Harrison:2002er} is 
given by a bimaximal neutrino mixing and an opportune form of the charged lepton 
mixing, as discussed in Ref.~\cite{Frampton:2004ud}. 
When trying to implement this scenario with a FN mechanism, one however encounters problems. Indeed, the assumption of 
bimaximal neutrino mixing poses strong 
constraints on the form of the light neutrino mass matrix, which can, most generally\footnote{Actually, there could be a 
second contribution that is exactly proportional to a unit matrix. Such a contribution, however, could never be explained 
by the FN mechanism, which can only explain orders of magnitude, but no exact equalities. Accordingly, this even more 
general case is not of practical relevance for FN inspired models.}, look 
like~\cite{Petcov:2004rk}:
\begin{equation}
 M_\nu^{\rm bi-max}= \begin{pmatrix}
 0 & A & B\\
 \bullet & 0 & 0\\
 \bullet & \bullet & 0
 \end{pmatrix}.
 \label{eq:Mnu_BM}
\end{equation}
Comparing this form with Eqs.~\eqref{eq:typeI} and~\eqref{eq:typeII}, 
we can see that the FN charges $(f_1,f_2,f_3)$ of the 
left-handed leptons need to be such that $|f_1+f_{2,3}|$ is small, while all other 
combinations, and in particular $|2 f_i|$, have to be large. This enforces large and more or 
less equal absolute values of all $f_i$'s, while simultaneously ${\rm sign}(f_1)=-{\rm sign}(f_{2,3})$ 
must be fulfilled. This is a very specific choice that cannot be brought in agreement 
with any of the possible forms of the charged lepton mass matrix $M_e$, listed in 
Ref.~\cite{Petcov:2004rk}, which could lead to a leptonic mixing in 
agreement with the experimental values. We have checked that these conclusions are 
not significantly altered by the use of two flavon fields.

%%%%%%%%%%%%%%%%%%%%%%%%%%%%%%%%%%%%%%%%%%%%%%%%%%%%%%%%%%%%%%%%%%%%%%
\subsection{\label{sec:LR} No Left-Right Symmetry}
%%%%%%%%%%%%%%%%%%%%%%%%%%%%%%%%%%%%%%%%%%%%%%%%%%%%%%%%%%%%%%%%%%%%%%

As we have just seen, it is relatively easy to impose requirements on the FN charges 
that are too restrictive to be fulfilled. This problem originates from the fact that it is 
non-trivial to obtain large mixings, as required for leptons, from a (non-lopsided) FN framework~\cite{Sato:2000ff}. 
As already seen in Sec.~\ref{sec:FN}, the symmetric form of the Majorana mass matrix essentially leads to some 
cascade-like structure at best, which can nevertheless lead to tri-bimaximal leptonic 
mixing in case that also the charged lepton mass matrix has a cascade-like form~\cite{Haba:2008dp}. 
It is exactly this last requirement, however, that is spoiled by Left-Right ($LR$-) symmetry.

Imposing $LR$-symmetry and including Higgs triplets~\cite{Deshpande:1990ip} in order to 
accommodate for the type II seesaw situation that is required in that context~\cite{Bezrukov:2009th}, 
the most general leptonic mass matrices must have the forms
\begin{eqnarray}
 (m_D)_{ij} &=& v_1 f_{ij} + v_2 g_{ij},\nonumber\\
 (M_e)_{ij} &=& v_1 g_{ij} + v_2 f_{ij},\nonumber\\
 (m_L)_{ij} &=& \sqrt{2} v_L h_{ij},\ {\rm and}\nonumber\\
 (M_R)_{ij} &=& \sqrt{2} v_R h_{ij},
 \label{eq:LR-matrices}
\end{eqnarray}
where $v_{1,2}$ are the Higgs doublet and $v_{L,R}$ are the Higgs triplet VEVs. 
However, in such a model the right-handed charged leptons $\overline{e_{iR}}$ are grouped 
into doublets $\Psi_R^i$ of $SU(2)_R$ together with the right-handed neutrinos $\overline{N_{iR}}$ 
[and hence their FN charges must be equal, $(k_1,k_2,k_3)=(g_1,g_2,g_3)$], and the discrete $LR$-symmetry 
dictates the equality of the absolute values of the FN charges between the left- and right-handed doublets, 
$Q(\Psi_L^i)= Q(\Psi_R^i)$ [and hence $(f_1,f_2,f_3)=(g_1,g_2,g_3)$, if we consider only 
positive FN charges]~\cite{Chen:2003zv}. From these conditions, our two Scenarios~A and~B already 
fix the complete structure of the mass matrices. Hence, the most general forms for the 
charged lepton and for the light neutrino mass matrices in Scenarios~(A,B) are given by
\begin{equation}
 M_e\propto
 \begin{pmatrix}
 1\hfill \hfill & \lambda^3\hfill \hfill & \lambda^{3,4}\\
 \lambda^3\hfill \hfill & 1\hfill \hfill & \lambda^{0,1}\\
 \lambda^{3,4} & \lambda^{0,1} & 1\hfill \hfill
 \end{pmatrix}\ \ {\rm and}\ \ \
 m_L\propto
 \begin{pmatrix}
 \lambda^{6,8} & \lambda^{3,5} & \lambda^{3,4}\\
 \bullet & \lambda^{0,2} & \lambda^{0,1}\\
 \bullet & \bullet & 1\hfill \hfill
 \end{pmatrix}.
\end{equation}
Since $M_e$ is a Dirac-type mass matrix, and since the above form is dictated directly by the choice of 
$(g_1,g_2,g_3)$, we cannot avoid a large 11-element, which clearly spoils the demanded 
cascade structure.\footnote{Our statement, however, only holds with certainty in 
a FN framework, and in particular for Majorana neutrinos, i.e., the light neutrino 
mass matrix is demanded to have a symmetric structure. 
Note that in the framework of Ref.~\cite{Haba:2008dp}, the authors suggest 
$LR$-symmetry to lead simultaneously to cascade structures in $M_e$ and $m_\nu$, which can be fulfilled in a non-FN context. 
In the cases discussed here, however, this is not possible.} 
In particular, none of the assignments that we 
will use later in the working examples fulfills the condition $Q(\Psi_L^i)=Q(\Psi_R^i)$.

%%%%%%%%%%%%%%%%%%%%%%%%%%%%%%%%%%%%%%%%%%%%%%%%%%%%%%%%%%%%%%%%%%%%%%
\subsection{\label{sec:2FN} More than one FN field}
%%%%%%%%%%%%%%%%%%%%%%%%%%%%%%%%%%%%%%%%%%%%%%%%%%%%%%%%%%%%%%%%%%%%%%

It had been shown in Ref.~\cite{Choi:2001rm} that, in 
a seesaw framework, it is difficult to obtain a large mixing angle scenario for leptons with only a single 
$U(1)_{\rm FN}$, unless one relies on pseudo-Dirac scenarios, which essentially involves setting some 
elements of the mass matrices equal to zero~\cite{Wolfenstein:1981kw}.

A very interesting comparison between one or instead two flavon 
fields has been performed in Ref.~\cite{Kanemura:2007yy} for the quark sector in an $SU(5)$ 
Grand Unified Theory (GUT) inspired scenario, and it has been 
extended in Ref.~\cite{Kamikado:2008jx} to the lepton sector. These references agree that, in order to have models which 
can be treated 
analytically, only real entries in the mass matrices should be investigated. 
This, however, will not allow for $CP$ violation, since all phases could be trivially rotated away. 
A way to accommodate non-trivial $CP$ phases is to introduce two flavon fields 
$\Theta_{1,2}$ rather than only one, and to require the VEVs of these two fields to have a 
relative phase. In addition, $\Theta_{1,2}$ must have opposite charges under an auxiliary $Z_2$ parity in order for this 
phase to survive. A further problem of models with only one FN field is that they normally lead to small atmospheric 
neutrino mixing~\cite{Kamikado:2008jx}, and are thus incompatible with the data. 
For these reasons, we are going to analyze in Sec.~\ref{sec:SU(5)} the case of two flavon fields, 
instead of having only one.

%%%%%%%%%%%%%%%%%%%%%%%%%%%%%%%%%%%%%%%%%%%%%%%%%%%%%%%%%%%%%%%%%%%%%%
\subsection{\label{sec:anomaly} Anomaly cancellation in $\mathbf{SU(5)}$}
%%%%%%%%%%%%%%%%%%%%%%%%%%%%%%%%%%%%%%%%%%%%%%%%%%%%%%%%%%%%%%%%%%%%%%

The FN charge assignments that we are going to use are coming from an $SU(5)$ GUT scenario~\cite{Ross:1985ai}, as 
mentioned in Ref.~\cite{Kamikado:2008jx} where, however, no detailed explanation for this form was given. We will do this 
by having a closer look at the conditions that have to be fulfilled in order to guarantee the cancellation of dangerous 
anomalies. 

The charge assignments used by us can be easily understood by looking at the conditions that have to be 
fulfilled in order to guarantee the cancellation of anomalies. Such 
considerations are used, for example, in the models from Refs.~\cite{Irges:1998ax,Ibanez:1994ig}. 
Here we will follow the procedure outlined in Refs.~\cite{Kane:2005va,Jain:1994hd}. 
In an $SU(5)$ GUT model, the right-handed electron is situated together with 
the quark doublets and the right-handed up-like quarks in a $\Rep{10}$-representation, 
whereas the lepton doublet is grouped together with the right-handed down-type 
quarks in a $\overline{\Rep{5}}$-representation. 
We use the parametrizations of Ref.~\cite{Kane:2005va}, where the authors define the 
FN charges of the quark doublets, of the right-handed up quarks, and right-handed electrons, 
respectively, as 
\be
\sum_{i=1}^3 q_i=x+u\,, \qquad \sum_{i=1}^3 u_i=x+2 u\,, \qquad \sum_{i=1}^3 e_i=x\,,
\ee
while the lepton doublets and right-handed down quarks are 
\be
\sum_{i=1}^3 l_i=y\,, \qquad \sum_{i=1}^3 d_i= y+v\,.
\ee
Anticipating the Assignments~(1,2) to be used later on in Sec.~\ref{sec:SU(5)}, 
cf.\ Tab.~\ref{tab:table_assignments}, we can determine the parameters $x,y,u,v$ from 
the conditions
\begin{eqnarray}
 k_1 + k_2 + k_3=5 &=& x+u = x+2u = x,\nonumber\\
 f_1 + f_2 + f_3=(1,4) &=& y = y+v,
 \label{eq:anomaly_conditions_1}
\end{eqnarray}
which immediately lead to $u=v=0$ (as characteristic for assignments consistent 
with $SU(5)$~\cite{Kane:2005va}), and hence also to $x=5$ and $y=(1,4)$. Since we will 
consider a case in which the SM-like Higgs (or other Higgses) do not carry any FN charge, we also need to fulfill 
$0=-z=z+u+v$~\cite{Kane:2005va}, which is no problem if $z=0$. Then, we can easily satisfy 
the condition for anomaly cancellation:
\begin{equation}
 A_3=A_2=\frac{3}{5} A_1= \frac{1}{2} [3x+4u+y+v]=\frac{1}{2} 
[3 (k_1 + k_2 + k_3) + f_1 + f_2 + f_3]=(8, 9.5),
 \label{eq:anomaly_conditions_2}
\end{equation}
while the $A'_1$ vanishes for both assignments, as demanded: 
\be
A'_1=\sum_{i=1}^3 (-q^2_i+2 u^2_i-d^2_i+l^2_i-e^2_i)=0\,.
\ee
Note that the FN charges $(g_1,g_2,g_3)$ 
of the right-handed neutrinos do not appear in the conditions displayed in 
Eqs.~\eqref{eq:anomaly_conditions_1} and~\eqref{eq:anomaly_conditions_2}. 
They are, indeed, total singlets not only under 
the SM gauge group, but also under $SU(5)$, and hence they cannot contribute to any gauge 
anomaly. 
This is the key point to be able to freely implement our right-handed neutrino 
Scenarios~A and~B.

%%%%%%%%%%%%%%%%%%%%%%%%%%%%%%%%%%%%%%%%%%%%%%%%%%%%%%%%%%%%%%%%%%%%%%
\subsection{\label{sec:SO(10)} Difficulties with $\mathbf{SO(10)}$?}
%%%%%%%%%%%%%%%%%%%%%%%%%%%%%%%%%%%%%%%%%%%%%%%%%%%%%%%%%%%%%%%%%%%%%%

We have decided to analyze in detail an $SU(5)$ inspired model. 
The reason for this is not only that the right-handed 
FN charges $(g_1,g_2,g_3)$ drop out of the light neutrino mass matrix, as they would in any 
type I or II seesaw model, but also that they are essentially unconstrained, since the right-handed 
neutrino $\overline{N_i}$ is a singlet $\Rep{1}_i$ for each generation $i$ in 
$SU(5)$~\cite{Ross:1985ai}. In an $SO(10)$ GUT~\cite{Ross:1985ai}, instead, the 
right-handed neutrino $\overline{N_i}$ is part of the $\Rep{16}_i$ representation together with all quarks 
and leptons of generation $i$~\cite{Asaka:2003fp}. This requirement would constrain the right-handed neutrino FN charges 
strongly, and it is therefore not clear if it is possible to find a realistic setup that reproduces all data correctly, 
while at the same time keeping a strong mass splitting in the right-handed neutrino sector.

%%%%%%%%%%%%%%%%%%%%%%%%%%%%%%%%%%%%%%%%%%%%%%%%%%%%%%%%%%%%%%%%%%%%%%
\subsection{\label{sec:non-dem} Problems with democratic Yukawa matrices}
%%%%%%%%%%%%%%%%%%%%%%%%%%%%%%%%%%%%%%%%%%%%%%%%%%%%%%%%%%%%%%%%%%%%%%

From a FN model, we generally expect a waterfall structure in 
the charged lepton and in the neutrino mass matrices~\cite{Haba:2008dp}. 
Let us suppose that a specific FN model leads to the following matrices, which have just the form 
that we will also obtain for both our scenarios, cf.\ Sec.~\ref{sec:SU(5)}, apart from overall factors: 
\be
M_e^\dagger M_e\propto \left(
\begin{array}{lll}
 \lambda ^4 & \lambda ^3 & \lambda ^2 \\
 \lambda ^3 & \lambda ^2 & \lambda  \\
 \lambda ^2 & \lambda  & 1
\end{array}
\right)\,, \quad 
m_\nu \propto \left(
\begin{array}{lll}
 \lambda ^2 & \lambda  & \lambda  \\
 \lambda  & 1 & 0 \\
 \lambda  & 0 & 1
\end{array}
\right)\,.
\label{eq:matrix_democratic}
\ee
Then, these matrices will be diagonalized respectively by $U_e$ 
and by $U_\nu \simeq U_\lambda U_H$: 
\be
U_e \simeq \left(
\begin{array}{lll}
 1 & 0 & 0 \\
 0 & 1 & \lambda  \\
 0 & -\lambda  & 1
\end{array}
\right)\,, \quad 
U_\lambda \simeq \left(
\begin{array}{lll}
 1 & \lambda  & \lambda  \\
 -\lambda  & 1 & 0 \\
 -\lambda  & 0 & 1
\end{array}
\right)\,, \quad 
U_H \simeq \left(
\begin{array}{lll}
 0 & 0 & 1 \\
 0 & 1 & 0 \\
 1 & 0 & 0
\end{array}
\right)\,,
\ee
with $U_\lambda$ being the matrix that diagonalizes $m_\nu$, and 
$U_H$ the matrix that corrects the neutrino eigenvalues for inverted ordering. Note that the 
correction  $U_H$ is required due to the presence of two large and only one small mass eigenvalues, 
which cannot be realized in normal hierarchy. In this case the PMNS matrix, 
\be
U_{\rm PMNS}\simeq 
\left(
\begin{array}{lll}
 \lambda  & \lambda  & 1 \\
 \lambda  & 1 & -\lambda  \\
 1 & -\lambda  & -\lambda 
\end{array}
\right)\,,
\label{eq:PMNS_democratic_wrong}
\ee
has a form which is not consistent with the data~\cite{Schwetz:2011qt}, as can be easily 
seen by calculating the value of $\theta_{13}$. 
The key points of this result are a waterfall structure for the 
mass matrices and the presence of the matrix $U_H$, and 
thus of an inverted ordering scenario in the neutrino sector. 

This simple argument shows that in general we need to go beyond the 
democratic Yukawa coupling assumption in a FN scenario. There is an 
easy way to overcome this problem, namely by considering a slightly 
non-democratic Yukawa hypothesis. We will demonstrate this explicitly in Sec.~\ref{sec:non-democratic} 
by considering specific $SU(5)$ inspired models, 
while in Sec.~\ref{sec:democratic} we will analyze the democratic case and find indeed a PMNS matrix of the 
form of Eq.~\eqref{eq:PMNS_democratic_wrong}.

%%%%%%%%%%%%%%%%%%%%%%%%%%%%%%%%%%%%%%%%%%%%%%%%%%%%%%%%%%%%%%%%%%%%%%
\subsection{\label{sec:RGE} No need for RGE running}
%%%%%%%%%%%%%%%%%%%%%%%%%%%%%%%%%%%%%%%%%%%%%%%%%%%%%%%%%%%%%%%%%%%%%%

The mass matrices we obtain are actually only correct at a high energy scale, 
like the GUT scale, where the FN charges are imposed. Since, 
however, the neutrino observables are measured at a low energy scale, 
we have to evolve the neutrino masses and mixing parameters down 
to that scale by renormalization group equations (RGEs)~\cite{Antusch:2005gp,Mei:2005qp,Bergstrom:2010qb}, an effect that 
is often dubbed as ``running''. Sometimes, this step is not applied, and it is instead argued that this 
should be the reason why a certain model does not fit to the data~\cite{Kamikado:2008jx}, 
although this must not necessarily hold true. But in general the running 
has to be taken into account, as there is no way to avoid it. We will, 
however, show in the following that running effects, although present, are fully negligible in 
our case.

First, note that the running of the charged lepton Yukawa coupling matrix 
$Y_e$ and of the light neutrino mass matrix $m_\nu$ in a type I seesaw 
model are given by~\cite{Antusch:2005gp},
\begin{eqnarray}
 16 \pi^2 \frac{d Y_e}{d t} &=& Y_e (D_e Y_e^\dagger Y_e + D_\nu 
Y_D^\dagger Y_D) + \ ({\rm flavour\ diagonal}), \label{eq:RGEeqs}\\
 16 \pi^2 \frac{d m_\nu}{d t} &=& (C_e Y_e^\dagger Y_e + 
C_\nu Y_D^\dagger Y_D)^T m_\nu + m_\nu (C_e Y_e^\dagger Y_e + 
C_\nu Y_D^\dagger Y_D) + \ ({\rm flavour\ diagonal}),\nonumber
\end{eqnarray}
where $t=\ln (\mu/\mu_0)$, $\mu$ is the renormalization scale, and $\mu_0$ 
is the reference scale (e.g.\ the GUT scale) at which the input information 
is imposed. The coefficients $C_{e,\nu}$ and $D_{e,\nu}$ are numbers of 
$\mathcal{O}(1)$. The flavour diagonal terms are not displayed, since they 
could only lead to an overall rescaling which is not important for mass 
ratios, but they will not affect the mixing. Note that we have neglected 
subtleties like threshold effects as they are not relevant to our argumentation.

Now, since 
we are considering a very low-scale seesaw framework, where the right-handed 
neutrino mass scale could be as low as a few GeV, the seesaw formula 
$m_\nu=-m_D^T M_R^{-1} m_D$ together with the definition of the Dirac neutrino mass, $m_D=y_D v$, implies that 
the order of the Dirac Yukawa coupling must be as small as, e.g., 
$y_D\sim 10^{-5}$ for $m_\nu=1$~eV and  $M_R\sim 10$~GeV. The dominant entry 
in $Y_e$, however, is $y_\tau \sim 0.01$. Accordingly one can, in Eq.~\eqref{eq:RGEeqs}, 
completely neglect $Y_D$, and safely assume that the largest number on the 
right-hand side is about $0.01^2=0.0001$. Dividing this number by the loop factor 
$16\pi^2$ decreases it to a value that is even smaller than the electron Yukawa 
coupling. Accordingly, any flavour non-diagonal term on the right-hand sides of 
Eq.~\eqref{eq:RGEeqs} will be small. Hence, the only effect the running can have 
is an overall scaling.  Furthermore, due to the smallness of $Y_D$, one can also 
neglect the running of the right-handed neutrino mass matrix~\cite{Antusch:2005gp}.

We can show more explicitly that the mixing angles and phases do not undergo a 
considerable running. The correction to the mixing angles can be estimated to be 
at most about~\cite{Antusch:2005gp}
\begin{equation}
 \Delta \theta \sim \frac{1}{16\pi^2}\,y_\tau^2\,\xi\,\ln \left( \frac{\mu_0}{\mu} \right),
 \label{eq:theta_RGE}
\end{equation}
where $\xi$ is an enhancement or suppression factor that can be at most as large as 
$\xi \sim \frac{\Delta m^2_A}{\Delta m^2_\odot} \sim 25$. Hence, from Eq.~\eqref{eq:theta_RGE}, 
one can estimate the maximal correction to a mixing angle to be about 0.001, which is 
perfectly negligible. Similarly, the evolution of the phases is, for a general phase 
$\phi$, roughly given by~\cite{Mei:2005qp}
\begin{equation}
 \Delta \phi \sim \frac{1}{16\pi^2}\,y_\tau^2\,\frac{1}{\zeta_{ij}}\,\ln \left( 
\frac{\mu_0}{\mu} \right) \sim \frac{10^{-5}}{\zeta_{ij}} ,
 \label{eq:phi_RGE}
\end{equation}
where $\zeta_{ij}=\frac{m_i-m_j}{m_i+m_j}$ is a function of the light neutrino masses. 
Obviously, this correction to the phase is also completely negligible, unless strong degeneracies in 
the neutrino masses lead to a very small $\zeta_{ij}$. Such a situation, however, is practically impossible to 
achieve in FN models, 
since any sensible FN charge assignment will always 
introduce hierarchies rather than degeneracies, so that we do not have to consider the 
running of any phases.

Accordingly, the only effect that running could have is an overall scaling of the mass 
matrices. Such scalings are, however, implicitly included in the prefactors of our mass 
matrices and, in particular, they will cancel out in mass ratios or ratios of mass 
squares. Furthermore, the running will practically not affect any mixing angles or 
$CP$ phases. We have verified numerically for several examples that this is indeed the 
case. In fact, for Yukawa couplings that are not larger than about $y_\tau$, numerical 
computations show no sign of running in the mixing angles in the absence of extreme 
degeneracies, even if we run over several orders of magnitude in energy~\cite{Bergstrom:2010qb}.

%%%%%%%%%%%%%%%%%%%%%%%%%%%%%%%%%%%%%%%%%%%%%%%%%%%%%%%%%%%%%%%%%%%%%%
\subsection{\label{sec:LFV} Potential constraints from lepton flavour violation}
%%%%%%%%%%%%%%%%%%%%%%%%%%%%%%%%%%%%%%%%%%%%%%%%%%%%%%%%%%%%%%%%%%%%%%

There is also a (seemingly unrelated) problem we want to comment on, which is, from a theoretical 
point of view, not necessarily connected to our models, but which might nevertheless show up 
in practice. This is the generic phenomenon that theories beyond the SM do not easily respect flavour, 
and will hence tend to lead to lepton flavour violating (LFV) reactions~\cite{Lee:1977tib,Blum:2007he}.

This problem arises in our case because we rely on a GUT framework which, in turn, requires a 
unification of the SM gauge couplings. However, with only the SM particle content, this unification 
does not happen~\cite{Martin:1997ns}. In principle, one would not necessarily have to care about 
this problem in a FN framework, since the FN mechanism intrinsically involves the existence of a 
high energy sector that is not specified further, and which is assumed to have just the right 
properties as to make the FN mechanism work~\cite{Froggatt:1978nt}. The easy way to go would be 
to simply assume this high energy sector, which is present anyway, to also be responsible for 
gauge coupling unification.

However, in practice, one would like to have a high energy sector at hand that is specified, 
in order to make concrete predictions. One of the known ways to achieve gauge coupling unification 
is to assume the presence of supersymmetry (SUSY)~\cite{Martin:1997ns}, which is often considered 
when talking about GUTs. But the introduction of SUSY also leads to problems connected to this 
theory, the prime example being generically large LFV effects~\cite{Hisano:1995cp}.

These effects have been studied in the context of GUT-inspired FN models, like ours, in 
Ref.~\cite{Sato:2000ff}. Even if a flavour diagonal universal slepton mass $m_0$ is assumed 
at the GUT scale, as typical for models inspired by the minimal supergravity (mSUGRA) scenario, 
RGE running will lead to growing flavour-violating effects at low energies. These off-diagonal 
elements can be estimated to have at most the size of~\cite{Sato:2000ff,Hisano:1995cp}
\begin{equation}
 \left( \Delta m^2_{\tilde L} \right)_{ij} \sim -\frac{6+2 a_0^2}{16 \pi^2} y_D^2 m_S^2 
\ln \left( \frac{\mu_0}{\mu} \right) U_{ik} U_{jk},
 \label{eq:slepton_off}
\end{equation}
where $a_0$ is an $\mathcal{O}(1)$ constant, $y_D\sim 10^{-5}$ is the largest Dirac Yukawa 
coupling (cf.\ Sec.~\ref{sec:RGE}), $m_S$ is the universal scalar mass (taken to be some typical 
superparticle mass), and $U_{rs}$ are elements of the PMNS matrix. Then, the maximal (if not even 
overestimated) value for the branching ratio of $\mu\to e\gamma$ can only be
\begin{equation}
 {\rm Br}(\mu \to e\gamma) \simeq \frac{\alpha^3}{G_F^2} \frac{\left( 
\Delta m^2_{\tilde L} \right)^2_{e \mu}}{m_S^8}\lesssim \frac{10^{-16}}{\left( m_S [{\rm GeV}] \right)^4},
 \label{eq:meg_BR}
\end{equation}
where we have used $U_{ik} U_{jk}\sim 1$ and $\ln ( \mu_0 /\mu )\sim 10$. However, even for scalar 
masses of only a few 100~GeV, the branching ratio is well below the current 90\%~C.L.\ limit of 
the MEGA experiment, ${\rm Br}(\mu^+ \to e^+ \gamma)<1.2\cdot 10^{-11}$~\cite{Brooks:1999pu,Ahmed:2001eh}.\footnote{Note that 
the next generation experiment MEG currently provides a slightly worse value, 
${\rm Br}(\mu^+ \to e^+ \gamma)<2.8\cdot 10^{-11}$ at 90\%~C.L.~\cite{Natori:2011zz}, which is expected to improve considerably 
within the next few years.} Other channels for LFV are less constrained~\cite{Raidal:2008jk}, and they are not expected to yield 
more stringent limits.

In conclusion, LFV processes make no problems in our case, since the Dirac Yukawa coupling $y_D$ in 
the models under consideration is very small, which we could also have concluded directly from the 
fact that we can easily ignore any running, cf.\ Sec.~\ref{sec:RGE}. As long as one does not include 
yet another potentially problematic high energy sector, our models are safe from this side.

%%%%%%%%%%%%%%%%%%%%%%%%%%%%%%%%%%%%%%%%%%%%%%%%%%%%%%%%%%%%%%%%%%%%%%
\subsection{\label{sec:proton} Proton decay in $\mathbf{SU(5)}$ GUTs}
%%%%%%%%%%%%%%%%%%%%%%%%%%%%%%%%%%%%%%%%%%%%%%%%%%%%%%%%%%%%%%%%%%%%%%

There is one final remark that is important for GUT theories: In most extensions of the Standard Model strong constraints arise from 
the requirements of gauge coupling unification and of perturbativity, like 
it was discussed in Ref.~\cite{Kopp:2009xt}. 
In particular, the scale $M_{\rm GUT}$ at which the gauge coupling unification occurs 
is related to the proton life-time $\tau_p$ through dimension-six operators. 
In supersymmetric theories, however, the leading contributions to 
proton decay come from dimension-five operators resulting from the
exchange of coloured higgsinos and winos. 
These diagrams produce an extremely fast proton decay and they 
have been used to constraint SUSY GUT models.

The supersymmetric $SU(5)$ in its minimal version has been tightly constrained 
(if not even ruled out) in Ref.~\cite{Murayama:2001ur} assuming that the gauge coupling 
unification is satisfied 
and imposing the limits provided by the Super-Kamiokande detector, which are particularly 
strong for the $p \rightarrow K^+ \bar{\nu}$ ($\tau_p >$~6.7~$\times$~10$^{32}$~years at 
90\%~C.L.~\cite{Kobayashi:2005pe}) and $p \rightarrow e^+ \pi^0$ channels ($\tau_p >$~8.2~$\times$~10$^{33}$~years 
at 90\%~C.L.~\cite{:2009gd}).

It is important, however, to note that even if the minimal SUSY $SU(5)$ is highly disfavoured, this does 
not mean that all possible non-minimal models are excluded as well, and some of them have ways to cure the 
proton decay problem: There exist several works in which non-minimal SUSY $SU(5)$ models have been vastly 
analyzed with the purpose of solving the proton decay problem. A possible way out is achieved by a more 
elaborate Higgs sector such that the mass of the Higgs triplet can be pushed to very heavy values, suppressing 
in this way dimension five operators. This is the so-called \emph{doublet-triplet splitting problem}, see, e.g., Ref.~\cite{Randall:1995sh} for a review. 
In this framework, however, it can be difficult to suppress proton decay and at the same time not to spoil gauge coupling 
unification. Other ways to suppress or eliminate completely dimension-five operators have been analyzed in 
the context of extra-dimension models, see Refs.~\cite{Nomura:2001mf,Kawamura:2000ev,Hall:2002ci} 
for five-dimensional SUSY $SU(5)$ theories, and in the context of a flipped 
$SU(5)$ model, where the up and down Yukawa couplings are reversed with respect to the standard $SU(5)$, see Ref.~\cite{Ellis:1995at}. Flavour symmetries have also been used to suppress dangerous proton decay 
operators~\cite{Murayama:1994tc}. Moreover, we want to remind the reader that the proton can naturally become
almost stable if $M_{\rm GUT}$ does not concide with the gauge coupling unification scale~\cite{Kawamura:2009re}. 
Finally, small group theoretical factors can help~\cite{Babu:1995cw}.

%%%%%%%%%%%%%%%%%%%%%%%%%%%%%%%%%%%%%%%%%%%%%%%%%%%%%%%%%%%%%%%%%%%%%%
\section{\label{sec:SU(5)} $\mathbf{SU(5)}$ inspired models with two FN fields}
%%%%%%%%%%%%%%%%%%%%%%%%%%%%%%%%%%%%%%%%%%%%%%%%%%%%%%%%%%%%%%%%%%%%%%

We now want to give explicit working examples that yield a simultaneous explanation of the 
low-energy neutrino and charged lepton data, as well as a working scenario with one keV-mass 
sterile neutrino. To do so, we start with an extension of the model presented in Ref.~\cite{Kamikado:2008jx}, 
which was based on Ref.~\cite{Kanemura:2007yy} and yields reasonable agreement with data. We 
will then relax three of their assumptions, in order to be able to construct fully working models: We will 
modify the charge assignments for the right-handed neutrinos, the ones for the left-handed lepton 
doublets, and we will depart from the fully democratic structure of the Yukawa matrices. Furthermore, we will present 
explicit analytical and numerical results in Secs.~\ref{sec:approx} and~\ref{sec:numerical}.

Let us start by recalling the ingredients taken from Ref.~\cite{Kamikado:2008jx}: First of all, 
the model needs two FN flavon fields $\Theta_{1,2}$ which obtain complex VEVs. Physically, we 
can always choose one VEV to be real, in our case $\langle \Theta_1 \rangle$. Then, the decisive 
quantities are the ratio $\lambda$ between the VEV $\langle \Theta_1 \rangle$ and the high-energy 
scale $\Lambda$, as well as the (complex) ratio $R$ between the VEVs:
\begin{equation}
 \lambda=\frac{\langle \Theta_1 \rangle}{\Lambda},\ \ R=\frac{\langle \Theta_1 \rangle}{\langle \Theta_2 \rangle} = 
R_0 e^{i\alpha_0},
 \label{eq:lambda_R}
\end{equation}
where $R_0$ and $\alpha_0$ are real numbers. Furthermore, as explained in Ref.~\cite{Kanemura:2007yy}, 
it is also necessary to introduce an auxiliary $Z_2$ symmetry, in order for the phase $\alpha_0$ in Eq.~\eqref{eq:lambda_R} 
to finally
be responsible for $CP$ violation. Next, we adopt the following FN charge and $Z_2$ assignments inspired by 
Refs.~\cite{Asaka:2003fp} and~\cite{Kanemura:2007yy}, respectively, for the FN fields and the lepton doublets, as well as for 
the right-handed charged leptons and neutrinos:
\begin{eqnarray}
 \Theta_{1,2}: && (-1,-1;+,-),\nonumber\\
 L_{1,2,3}: && (a+1,a,a;+,+,-),\nonumber\\
 \overline{e_{1,2,3}}: && (3,2,0;+,+,-),\nonumber\\
 \overline{N_{1,2,3}}: && (g_1,g_2, g_3; +,+,-),
 \label{eq:FN-assignments}
\end{eqnarray}
where $a=0,1$ (see Tab.~\ref{tab:table_assignments}). 
In general, we denote the FN charges of the flavon fields by $(\theta_1,\theta_2)$, the ones of the lepton doublets 
by $f_i$, and the ones of the right-handed charged leptons by $k_i$, as in Sec.~\ref{sec:FN}. The charges $g_i$ 
of the right-handed neutrinos will be chosen according to our Scenarios~A and~B introduced in Sec.~\ref{sec:MR}. 
The key point is that the FN charges of the right-handed neutrinos will drop out of the light neutrino mass matrix 
when constructing the light neutrino mass matrix using the seesaw mechanism of type~I or~II, as explained in 
Refs.~\cite{Kamikado:2008jx,Choi:2001rm,Kanemura:2007yy}. This allows us to freely choose $g_i$ without changing the appearance 
of the light 
neutrino mass matrix. It may be, however, that the charges $g_i$ are constrained by other consistency 
conditions (cf.\ Sec.~\ref{sec:SO(10)}).

\begin{table}[t]
 \centering
 \begin{tabular}{|c||c|c|c|}\hline
 Field & $L_1$ & $L_2$ & $L_3$ \\\hline\hline
 Assignment 1 ($a=0$) & $(1,+)$ & $(0,+)$ & $(0,-)$\\ \hline
 Assignment 2 ($a=1$) & $(2,+)$ & $(1,+)$ & $(1,-)$\\ \hline
 \end{tabular}
 \caption{\label{tab:table_assignments} Family and $Z_2$ charges for the lepton doublets.}
\end{table}

Before proceeding, we also want to show how the mass matrices are constructed. 
The most general (seesaw type II) Lagrangian that leads to masses in the lepton sector is
\begin{eqnarray}
 \mathcal{L} &=& 
-\sum_{a,b,i,j}^{a+b=k_i+f_j} Y_e^{ij}\,\overline{e_{iR}}\,H\,L_{jL}\,\lambda_1^a \lambda_2^b +h.c.\
-\sum_{a,b,i,j}^{a+b=g_i+f_j} Y_D^{ij}\,\overline{N_{iR}}\,\tilde{H}\,L_{jL}\,\lambda_1^a \lambda_2^b +h.c. \\
 && - \sum_{a,b,i,j}^{a+b=f_i+f_j} \frac{1}{2} \overline{(L_{iL})^C}\,\tilde m_L^{ij}\,L_{jL}\,\lambda_1^a \lambda_2^b +h.c.\
- \sum_{a,b,i,j}^{a+b=g_i+g_j} \frac{1}{2}\overline{N_{iR}}\,\tilde M_R^{ij}\,(N_{jR})^C\,\lambda_1^a \lambda_2^b+h.c.\,,
\nonumber
 \label{eq:FN-Lagrangian}
\end{eqnarray}
where, by using Eq.~\eqref{eq:lambda_R},
\begin{equation}
 \lambda_1^a \lambda_2^b \equiv \left( \frac{\Theta_1}{\Lambda} \right)^a \left( \frac{\Theta_2}{\Lambda} \right)^b = 
\lambda^{a+b} R^b.
 \label{eq:L-ratios}
\end{equation}
The matrices $Y_e$, $Y_D$, and $\tilde M_R$ are the charged lepton Yukawa matrix, the Dirac 
neutrino Yukawa matrix, and the uncorrected right-handed Majorana neutrino mass matrix, respectively, which 
where all taken in Ref.~\cite{Kamikado:2008jx} to have a democratic structure. In addition to 
that, $\tilde m_L$ is the uncorrected left-handed Majorana neutrino mass matrix, which is present in type II 
seesaw cases. Note that the matrix elements of $\tilde M_R$ and $\tilde m_L$ are all of the same 
order, i.e., not yet corrected by FN contributions. The sums run over all possible 
values of $a$ and $b$ that fulfill the condition of full cancellation of the FN charges in each term. 
Furthermore, certain terms may violate the $Z_2$ parity and must hence be set to zero.

Using all this, we can derive the mass matrices for four different cases: Assignment~1 or~2, each 
combined with Scenario~A or~B (for type~I or type~II seesaw), respectively, always separated by commas in the respective equations. For the charged leptons, we obtain
\begin{equation}
 M_e^{(1,2)} = v \begin{pmatrix}
 Y_e^{11} B_{2,4} \lambda^{3,4}\hfill \hfill & Y_e^{12} B_2 \lambda^{2,3}\hfill \hfill & Y_e^{13} B_{0,2} R \lambda^{2,3}
\hfill \hfill\\
 Y_e^{21} B_2 \lambda^{2,3}\hfill \hfill & Y_e^{22} B_{0,2} \lambda^{1,2}\hfill \hfill & Y_e^{23} R \lambda^{1,2}\hfill \hfill\\
 Y_e^{31} R \lambda^{1,2} \hfill \hfill & 0,\ Y_e^{32} R \lambda \hfill \hfill  & Y_e^{33} \lambda^{0,1} \hfill \hfill \hfill
 \end{pmatrix}.
 \label{eq:charged_matrices_12}
\end{equation}
The right-handed neutrino mass matrices for Scenarios~A and~B turn out to be
\begin{equation}
  M_R^{\rm (A,B)} = \begin{pmatrix}
 \tilde M_R^{11} B_{6,8} \lambda^{6,8} & \tilde M_R^{12} B_{2,4} \lambda^{3,5} & \tilde M_R^{13} R B_2 \lambda^{3,4}\\
 \bullet & \tilde M_R^{22} B_{0,2} \lambda^{0,2} & 0,\ \tilde M_R^{23} R \lambda \hfill \hfill \\
 \bullet & \bullet  & \tilde M_R^{33}\hfill \hfill
 \end{pmatrix},
 \label{eq:MR_matrices_AB}
\end{equation}
while the left-handed ones for Assignments~1 and~2 are given by
\begin{equation}
  m_L^{(1,2)} = \begin{pmatrix}
 \tilde m_L^{11} B_{2,4} \lambda^{2,4}\hfill \hfill & \tilde m_L^{12} B_{0,2} \lambda^{1,3}\hfill \hfill & \tilde 
m_L^{13} R B_{0,2} \lambda^{1,3}\hfill \hfill\\
 \bullet & \tilde m_L^{22} B_{0,2} \lambda^{0,2}\hfill \hfill & 0,\ \tilde m_L^{23} R \lambda^2\hfill \hfill \\
 \bullet & \bullet  & \tilde m_L^{33} B_{0,2} \lambda^{0,2}\hfill \hfill
 \end{pmatrix}.
 \label{eq:mL_matrices_12}
\end{equation}
Finally, the Dirac mass matrices for Assignment~1 are
\begin{equation}
  m_D^{\rm (1A,1B)} = v \begin{pmatrix}
 Y_D^{11} B_4 \lambda^{4,5} \hfill \hfill & Y_D^{12} B_{2,4} \lambda^{3,4}\hfill \hfill & Y_D^{13} R B_2 \lambda^{3,4} 
\hfill \hfill \\
 Y_D^{21} B_{0,2} \lambda^{1,2}\hfill \hfill & Y_D^{22} \lambda^{0,1} \hfill \hfill & 0,\ Y_D^{23} R \lambda \hfill \hfill \\
 Y_D^{31} R \lambda \hfill \hfill & 0 \hfill \hfill  & Y_D^{33} \hfill \hfill
 \end{pmatrix},
 \label{eq:mD_matrices_1AB}
\end{equation}
while the ones for Assignment~2 turn out to be
\begin{equation}
  m_D^{\rm (2A,2B)} = v \begin{pmatrix}
 Y_D^{11} B_{4,6} \lambda^{5,6} \hfill \hfill & Y_D^{12} B_4 \lambda^{4,5}\hfill \hfill & Y_D^{13} R B_{2,4} \lambda^{4,5} 
\hfill \hfill \\
 Y_D^{21} B_2 \lambda^{2,3}\hfill \hfill & Y_D^{22} B_{0,2} \lambda^{1,2} \hfill \hfill & Y_D^{23} R \lambda^{1,2} 
\hfill \hfill \\
 Y_D^{31} R \lambda^2 \hfill \hfill & Y_D^{32} R \lambda \hfill \hfill  & Y_D^{33} \lambda \hfill \hfill
 \end{pmatrix},
 \label{eq:mD_matrices_2AB}
\end{equation}
where $v=174$~GeV is the electroweak VEV, and $B_{2n}=1+R^2+...+R^{2n}$. Note that, due to $g_{2,3}=0$ in Scenario~A, the 
23-entry of $M_R$ is actually forbidden by the $Z_2$ parity in that case.

From $M_R$ in Eq.~\eqref{eq:MR_matrices_AB}, one can immediately calculate the mass eigenvalues 
for the right-handed neutrinos as functions of the right-handed mass scale $M_0$:
\begin{eqnarray}
 {\rm A}(3,0,0): && M_1 = M_0 \lambda^6\ 2 R_0^2 \sqrt{1+ R_0^4 + 2 R_0^2 \cos (2\alpha_0)}\,, \nonumber\\
                 && M_2 = M_0\,, \nonumber\\
                 && M_3 = M_0 \left(1 + \lambda^6 [ 1 + R_0^2 (3 \cos (2 \alpha_0) + 3 R_0^2 \cos (4\alpha_0) + 
R_0^4 \cos (6 \alpha_0) ] \right)\,, \nonumber\\
 {\rm B}(4,1,0): && M_1 = M_0 \lambda^8\ 2 R_0^4 \sqrt{1 + R_0^8 - 2 R_0^4 \cos (4 \alpha_0)}\,,\nonumber\\
                 && M_2 = M_0 \lambda^2\,, \nonumber\\
                 && M_3 = M_0 \ \left( 1+ R_0^2 \lambda^2 \cos (2\alpha_0) \right)\,. \nonumber
\end{eqnarray}
Indeed, the eigenvalues of $M_R$ show just the hierarchical structure that we have expected: 
If we suppose that $M_1$ is of $\mathcal{O}({\rm keV})$, then $M_0$ should be about $10^6~{\rm keV}\sim 1~{\rm GeV}$ for 
Scenario~A, or about $10^8~{\rm keV}\sim 100~{\rm GeV}$ for Scenario~B. In any case, we would have a low-scale seesaw to work. 
One can of course raise the possible value for $M_0$ by simply assigning an even higher charge $g_1$ to the first generation 
right-handed neutrino: Already for $g_1=5$, one could have $M_0\sim 10$~TeV, while one would only have to be careful to keep 
$g_1-g_2\geq 3$, as explained in Sec.~\ref{sec:MR}.

%%%%%%%%%%%%%%%%%%%%%%%%%%%%%%%%%%%%%%%%%%%%%%%%%%%%%%%%%%%%%%%%%%%%%%
\subsection{\label{sec:approx} Analytical results}
%%%%%%%%%%%%%%%%%%%%%%%%%%%%%%%%%%%%%%%%%%%%%%%%%%%%%%%%%%%%%%%%%%%%%%

In this section, we will exemplify for Assignment~1 (and seesaw type~I), how one can arrive at analytical 
approximations for the masses 
and for the PMNS-matrix. We will also see explicitly that it is necessary to depart from the 
democratic structure of the Yukawa matrices in order to obtain sensible results (cf.\ Sec.~\ref{sec:non-dem}), i.e., 
the FN mechanism alone is not strong enough to fully explain the data. It will turn out, however, that this simple 
analytical consideration is still not perfect, even in the non-democratic case, and we have to rely on numerics in order to find 
quasi-perfect models, 
which will be discussed in Sec.~\ref{sec:numerical}.

%%%%%%%%%%%%%%%%%%%%%%%%%%%%%%%%%%%%%%%%%%%%%%%%%%%%%%%%%%%%%%%%%%%%%%
\subsubsection{\label{sec:democratic} Democratic matrices}
%%%%%%%%%%%%%%%%%%%%%%%%%%%%%%%%%%%%%%%%%%%%%%%%%%%%%%%%%%%%%%%%%%%%%%

The natural starting point is having democratic forms of all Yukawa and bare mass matrices, i.e.,
\begin{equation}
 Y_e^{ij}=y_e\,,\; Y_D^{ij}=y_D\,,\; {\rm and}\; \tilde M_R^{ij}=M_0\;\; \forall i,j.
 \label{eq:democratic}
\end{equation}
In this case, the type~I light neutrino mass matrices can be easily calculated from Eqs.~\eqref{eq:MR_matrices_AB} 
and~\eqref{eq:mD_matrices_1AB}, using the seesaw formula $m_\nu=-m_D^T M_R^{-1} m_D$. For Scenario~A, this 
results in
\begin{equation}
  m_\nu^{\rm (1A), I} = m_{\nu 0} \begin{pmatrix}
  \frac{ \lambda ^2 \left(-2 B_2 B_4 \left(R^2+1\right)+B_4^2+B_6 R^2+B_6\right)}{B_2^2
   \left(R^2+1\right)-B_6} & - \lambda  & - R \lambda  \\
 - \lambda  & -1 & 0 \\
 - R \lambda  & 0 & -1
 \end{pmatrix},
 \label{eq:mNu_matrices_I1A}
\end{equation}
whereas the corresponding expression for Scenario~B is given by
\begin{equation}
  m_\nu^{\rm (1B), I} = m_{\nu 0} \begin{pmatrix}
  -B_2 \lambda ^2 & - \lambda  & - R \lambda  \\
 - \lambda  & \frac{ \left(-B_2^2 R^2+B_4^2 \left(B_2-R^2-2\right)+2 B_2 B_4
   R^2+B_8\right)}{B_2^3 R^2-2 B_2 B_4 R^2-B_2 B_8+B_4^2+B_8 R^2} & 0 \\
 - R \lambda  & 0 & -1
 \end{pmatrix},
 \label{eq:mNu_matrices_I1B}
\end{equation}
where $m_{\nu 0}=\frac{y_D^2 v^2}{M_0}$ in both cases. In order to obtain the corresponding mixing, we need to diagonalize 
the two neutrino mass matrices in Eqs.~\eqref{eq:mNu_matrices_I1A} and~\eqref{eq:mNu_matrices_I1B}, 
as well as the charged lepton mass matrix for Assignment~1 from Eq.~\eqref{eq:charged_matrices_12}. 
It turns out that all these matrices can be approximately diagonalized by applying a series of small and large rotation 
matrices, all of which are (at least approximately) unitary. Such a stepwise diagonalization, although certainly not suitable 
for general mass matrices, is ideally suited for FN models, since the corresponding matrices intrinsically involve the small 
parameter $\lambda$ in which all rotation matrices can be Taylor expanded. Note that, in order to derive the leptonic mixing 
matrices, we follow the conventions used in the {\it Mixing Parameter Tools (MPT)} 
package~\cite{Antusch:2005gp,MPT}, which will also be used later on for our numerical computations.

In order to obtain the (unitary) charged lepton mixing matrix $U_e$, we have to diagonalize the squared charged lepton 
mass matrix $M_e^\dagger M_e$ by
\begin{equation}
 U_e^\dagger M_e^\dagger M_e U_e= {\rm diag} (m_e^2, m_\mu^2, m_\tau^2).
 \label{eq:CL_diagonalization}
\end{equation}
In our case, Eq.~\eqref{eq:charged_matrices_12}, it turns out that the charged lepton mixing matrix is given by
\begin{equation}
 U_e= U_A U_B U_C U_D U_E U_F U_G,
 \label{eq:CL_mixing}
\end{equation}
where the respective matrices are all at least approximately unitary and are reported in Eq.~\eqref{eq:CL_Umatrices}. 
These subsequent transformations bring $M_e^\dagger M_e$ to an approximately diagonal form, from which one can read off 
the expressions for the charged lepton masses to be
\be
\left\{
\begin{array}{lcl}
 m_e &=& m_0 \lambda ^3 \; R_0^2\,,\nonumber\\
 m_\mu &=& m_0 \lambda \; \left( 1+ \lambda^2 \left[ R_0^2 \cos (2 \alpha_0 )+\frac{R_0^4-R_0^2+3}{2} \right] \right)\,,
\nonumber\\
 m_\tau &=& m_0 \; \left( 1 + \frac{3}{2} R_0^2 \lambda^2 \right)\,.
 \label{eq:CL_masses}
\end{array}
\right.
\ee
From these relations, we can determine the mass ratios: $m_e/m_\mu\simeq R_0^2 \lambda^2$ and 
$m_\mu/m_\tau\simeq\lambda$. Using the measured values of $m_e$, $m_\mu$, and $m_\tau$, we find 
the sizes of the parameters to be roughly
\begin{equation}
 \lambda \simeq 0.06 \quad {\rm and}\quad R_0\simeq 1.18\,.
 \label{eq:CL_ratios}
\end{equation}
As noted in Ref.~\cite{Sato:2000ff}, the most advantageous choice of the parameter $\lambda$ turns out to be a bit below the 
standard choice $0.22$. The phase $\alpha_0$ is not fixed by the lowest order expressions in Eq.~\eqref{eq:CL_ratios}, but one 
can choose it to have the value $\alpha_0=0.67$ in order to make the $\mathcal{O}(\lambda^3)$-correction to $m_\mu/m_\tau$ 
vanish.\footnote{For Assignment~2, we would have obtained $m_e/m_\mu\simeq R_0^4 \sqrt{1+R_0^2} \lambda^2$ and 
$m_\mu/m_\tau\simeq \frac{\lambda}{1+R_0^2}$, leading to $\lambda\simeq 0.10$ and $R_0\simeq 0.80$. The phase $\alpha_0$ is 
unconstrained, but the choice $\alpha_0=0.60$ turns out to be numerically convenient.}

Let us now have a look at the neutrino mass matrix for Scenario~A. The matrix $m_\nu^{\rm (1A), \rm I}$ 
in Eq.~\eqref{eq:mNu_matrices_I1A} can be diagonalized by a unitary matrix $U_\nu\equiv U_\nu^{\rm (1A), I}$ 
according to $U_{\nu}^T m_\nu^{(1A), \rm I} U_{\nu}={\rm diag}(m_1,m_2,m_3)$, where the mass eigenvalues 
$m_i$ can still contain complex phases. Also this can be done by a stepwise diagonalization resulting in
\begin{equation}
 U_\nu= U_\alpha U_\beta U_\gamma U_\delta U_\epsilon U_\zeta U_\eta\,,
 \label{eq:Nu_mixing_1A}
\end{equation}
with the respective pieces being all at least approximately unitary and are given by Eq.~\eqref{eq:Nu_Umatrices_1A}. 
Note that the purpose of the last matrix $U_\eta$ is only to reshuffle the mass eigenvalues in order to accommodate 
for the correct ordering, since the resulting pattern (two larger eigenvalues and one smaller one) can only be realized in 
inverted ordering. It is exactly this point, which will change in the non-democratic cases, and this 
is also the reason why the democratic cases provide a worse match to the data. The neutrino mass eigenvalues can be 
determined as the absolute values of the diagonal entries of the resulting mass matrix. They turn out to be
\be
\left\{
\begin{array}{lcl}
 |m_1| &=& m_{\nu 0}\;[1+ \lambda^2 (1+R_0^2-\sqrt{(-1+R_0^2)^2+R_0^2 \cos^2 \alpha_0 })]+\mathcal{O}(\lambda^4)\,,\nonumber\\
 |m_2| &=& m_{\nu 0}\;[1+ \lambda^2 (1+R_0^2+\sqrt{(-1+R_0^2)^2+R_0^2 \cos^2 \alpha_0 })]+\mathcal{O}(\lambda^4)\,,\nonumber\\ 
 |m_3| &=& m_{\nu 0}\lambda ^2\; \frac{R_0^2 }{2 \sqrt{R_0^4+2 R_0^2 \cos (2 \alpha_0)+1}}+\mathcal{O}(\lambda^4)\,.
 \label{eq:Nu_eigenvals_1A}
\end{array}
\right.
\ee
The mass square differences are, at lowest order,
\be
\left\{
\begin{array}{lcl}
 \Delta m_\odot^2 &\simeq& 4 m_{\nu 0}^2 \lambda^2 \; \sqrt{(-1+R_0^2)^2+R_0^2\cos^2 \alpha_0 } +\mathcal{O}(\lambda^4)\,,
\nonumber\\ 
 \Delta m_A^2 &\simeq& m_{\nu 0}^2 \;(1+2 \lambda^2 [ 1+R_0^2-\sqrt{(-1+R_0^2)^2+R_0^2\cos^2 \alpha_0 }] +\mathcal{O}(\lambda^4)\,.
 \label{eq:Nu_Deltas_1A}
\end{array}
\right.
\ee
Using the numbers from Eq.~\eqref{eq:CL_ratios}, one can predict the ratio 
$\frac{\Delta m_\odot^2}{\Delta m_A^2}\simeq [0.018,0.005]$, respectively 
for $\alpha_0=[0,\pi/2]$, where the choice $\alpha_0=0.67$ leads to the value 0.015. To obtain a fair agreement with the 
measured value of about 0.031 
we would need $\alpha_0=0$ and thus no $CP$ violation. 
In addition to the imperfect prediction for $\frac{\Delta m_\odot^2}{\Delta m_A^2}$, the democratic 
form of the Yukawa matrices does lead to a bad mixing. This fact is visible 
when looking at the full PMNS-matrix, $U_{\rm PMNS}=U_e^\dagger U_\nu$, which is given by
\begin{equation}
 U_{\rm PMNS}^{\rm (1A), I}=
\left(
\begin{array}{lll}
  0 & 0 & 1+\left(R_0^2+1\right) \lambda ^2\\
  \frac{-\zeta_1}{\sqrt{R_0^2+\zeta_1^2}} + u^{\rm (1A),I}_{21} \lambda^2  
& \frac{\zeta_1}{\sqrt{R_0^2+\zeta_1^2}} + u^{\rm (1A),I}_{21} \lambda^2 & 0 \\
  \frac{R_0}{\sqrt{R_0^2+\zeta_1^2}} + u^{\rm (1A),I}_{31} \lambda^2 
& \frac{R_0}{\sqrt{R_0^2+\zeta_1^2}} + u^{\rm (1A),I}_{32} \lambda^2 & 0
 \label{eq:PMNS_1A}
\end{array}
\right)
 + \mathcal{O}(\lambda^3)\,,
\end{equation}
where
\begin{eqnarray}
 && u^{\rm (1A),I}_{21}=\frac{-R_0^2 (2 \cos \alpha_0+i \sin \alpha_0 -2 \zeta_1)}{2 \sqrt{R_0^2+\zeta_1^2}},\ 
u^{\rm (1A),I}_{31}= \frac{R_0 \left(2 R_0^2 +\zeta_1 [i \sin \alpha_0 -2 \cos  \alpha_0 ]\right)}{2 \sqrt{R_0^2+\zeta_1^2}},
\nonumber\\ 
 && u^{\rm (1A),I}_{32}= \frac{R_0 \left(2 R_0^2 +\zeta_1 [2 \cos \alpha_0 -i \sin \alpha_0 ]\right)}{2 \sqrt{R_0^2+\zeta_1^2}},
\label{eq:u_dem}
\end{eqnarray}
and $\zeta_1$ is defined in Eq.~\eqref{eq:Nu_entries_1A}. 
Obviously, this matrix does not have the desired form, since the angle $\theta_{13}$ is extremely close to the maximal 
value $\pi/2$, instead of being small or even vanishing.

Let us now check if Scenario~B can give a better match with data. Here, the unitary matrix $U_\nu\equiv U_\nu^{\rm (1B), I}$ 
is given by
\begin{equation}
 U_\nu= U'_\alpha U'_\beta U'_\gamma U'_\delta U'_\epsilon U'_\zeta\,,
 \label{eq:Nu_mixing_1B}
\end{equation}
where the individual rotation matrices are reported in Eq.~\eqref{eq:Nu_Umatrices_1B}. 
Again, the last matrix $U'_\zeta$ corrects for inverted ordering. The eigenvalues read
\be
\left\{
\begin{array}{lcl}
 |m_1| &=& m_{\nu 0}\; (1+2 R_0^2 \lambda^2)+ \mathcal{O}(\lambda^4)\,,\nonumber\\ 
 |m_2| &=& m_{\nu 0}\; \frac{\left(R_0^4 \left(\lambda^2+2\right)-2 R_0^2 \lambda^2 \cos (2 \alpha_0 )
+\lambda^2\right)}{R_0^2 \sqrt{R_0^4-2 R_0^2 \cos (2 \alpha_0 )+1}}+\mathcal{O}(\lambda^4)\,,\nonumber\\
 |m_3| &=& m_{\nu 0} \lambda^2 \; \frac{\sqrt{R_0^4+2 R_0^2 \cos (2 \alpha_0 )+1}}{2 R_0^2}+ \mathcal{O}(\lambda^4)\,,
 \label{eq:Nu_eigenvals_1B}
\end{array}
\right.
\ee
and the mass square differences are, at lowest order,
\be
\left\{
\begin{array}{lcl}
 \Delta m_\odot^2 &\simeq& m_{\nu 0}^2 \; \left(-4 \left(R_0^2-1\right) \lambda^2
+\frac{4 R_0^4}{R_0^4-2 R_0^2 \cos (2 \alpha_0 )+1}-1\right)+ \mathcal{O}(\lambda^4)\,,\nonumber\\
 \Delta m_A^2 &\simeq& m_{\nu 0}^2 \; (1+ 4 R_0^2 \lambda^2 )+ \mathcal{O}(\lambda^4)\,.
 \label{eq:Nu_Deltas_1B}
\end{array}
\right.
\ee
Then, even with the most suitable value of $\alpha_0=\frac{\pi}{2}$, the ratio of mass square 
differences, $\frac{\Delta m_\odot^2}{\Delta m_A^2}\simeq 0.35$, turns out to be much too large. 
Also the PMNS matrix,
\begin{equation}
 U_{\rm PMNS}^{\rm (1B), I}=\begin{pmatrix}
 0 & -\frac{i \left(R_0^2+e^{2 i \alpha_0 }\right)}{2 R_0^2} \lambda  
 & 1+\left(R_0^2+\frac{1}{2}-\frac{e^{-2 i \alpha_0 }}{2 R_0^2}\right)\lambda^2 \\
 u^{\rm (1B), I}_{21} \lambda^2  
 & \frac{1}{2} i \left(2+\left(1-\frac{e^{2 i \alpha_0 }}{R_0^2}\right)\lambda^2\right) 
 & \frac{1}{2} \left(1+\frac{e^{-2 i \alpha_0 }}{R_0^2}\right) \lambda  \\
 i \left(1+R_0^2 \lambda^2\right) 
 & \frac{2 i e^{-i \alpha_0 } R_0^3 \left(-e^{4 i \alpha_0 }+2 R_0^2+e^{2 i \alpha_0}R_0^2\right)}{3 R_0^4+2\cos(2 \alpha_0) 
R_0^2-1} \lambda^2 
 & 0
 \end{pmatrix}+ \mathcal{O}(\lambda^3)\,,
 \label{eq:PMNS_1B}
\end{equation}
where
\begin{equation}
 u^{\rm (1B), I}_{21} = -\frac{2 i e^{-3 i \alpha_0} R_0^3 
\left[-1 + e^{2i \alpha_0}(1+ 2 e^{2 i \alpha_0} R_0^2) \right]}
{3 R_0^4+2 \cos (2 \alpha_0 )R_0^2-1},
 \label{eq:PMNS_1B_aux}
\end{equation}
turns out not to be much better than the one from Eq.~\eqref{eq:PMNS_1A}. 
From the above calculations it is clear that the hypothesis of democratic Yukawa matrices 
does not lead to a satisfactory PMNS matrix. 
In the following section, we will analyze the non-democratic case to see if we might be able 
to achieve better agreement with the neutrino data.

%%%%%%%%%%%%%%%%%%%%%%%%%%%%%%%%%%%%%%%%%%%%%%%%%%%%%%%%%%%%%%%%%%%%%%
\subsubsection{\label{sec:non-democratic} Slightly non-democratic matrices}
%%%%%%%%%%%%%%%%%%%%%%%%%%%%%%%%%%%%%%%%%%%%%%%%%%%%%%%%%%%%%%%%%%%%%%

The problem of fully democratic matrices not being perfectly suitable for $SU(5)$ inspired FN models has already been 
discussed in Ref.~\cite{Sato:2000kj}. The key point to obtain a better agreement with data is to change the form of the 
light neutrino mass matrix. Ideally, they should have the form
\begin{equation}
m_\nu \propto
 \begin{pmatrix}
 \lambda^2 & \lambda & \lambda\\
 \lambda &\delta_0^2 & 0\\
 \lambda & 0 & 1
 \end{pmatrix},
 \label{eq:23-ideal}
\end{equation}
with $\delta_0^2=\sqrt{\frac{\Delta m_\odot^2}{\Delta m_A^2}}\simeq 0.18>\lambda$, whereas our matrices in 
Eqs.~\eqref{eq:mNu_matrices_I1A} and~\eqref{eq:mNu_matrices_I1B} instead have the form
\begin{equation}
m_\nu \propto
 \begin{pmatrix}
 \lambda^2 & \lambda & \lambda\\
 \lambda & 1 & 0\\
 \lambda & 0 & 1
 \end{pmatrix},
 \label{eq:23-ours}
\end{equation}
which looks similar, but causes a major difference for the mass eigenvalues and for the mixings. 
A suitable modification can be obtained by simply choosing the Yukawa couplings $Y_D^{12}=Y_D^{22}=\delta_0 y_D$ instead 
of $Y_D^{12}=Y_D^{22}=y_D$. Indeed, due to $\delta_0\sim 0.30$, this is just a relatively mild deviation from the 
fully democratic case, and it can easily be justified by varying the entries in the Yukawa matrices in a certain range 
around their natural values~$y_D$.

Using this modification, the new light neutrino masses look like
\begin{equation}
 m_\nu^{\rm (1A), I}|_{\rm non-d.}= m_{\nu 0} \begin{pmatrix}
 \frac{\lambda^2 \left(-2 B_2 B_4 \left(R^2+1\right)+ B_4^2+B_6 R^2+ B_6\right)}{B_2^2
   \left(R^2+1\right)-B_6} & - \delta_0  \lambda  & - R \lambda  \\
 - \delta_0  \lambda  & - \delta_0^2 & 0 \\
 - R \lambda  & 0 & -1
 \end{pmatrix},
 \label{eq:mNu_matrices_I1A_nonD}
\end{equation}
\begin{equation}
 m_\nu^{\rm (1B), I}|_{\rm non-d.}= m_{\nu 0} \begin{pmatrix}
 -B_2 \lambda^2 & - \delta_0  \lambda  & - R \lambda  \\
 - \delta_0  \lambda  & \frac{ \delta_0^2 \left(-B_2^2 R^2+B_4^2 \left(B_2-R^2-2\right)+2 B_2
   B_4 R^2+B_8\right)}{B_2^3 R^2-2 B_2 B_4 R^2-B_2 B_8+B_4^2+B_8 R^2} & 0 \\
 - R \lambda  & 0 & -1
 \end{pmatrix}.
 \label{eq:mNu_matrices_I1B_nonD}
\end{equation}
The matrix in Eq.~\eqref{eq:mNu_matrices_I1A_nonD} can again be diagonalized by a unitary matrix $U_\nu$, 
which is now given by
\begin{equation}
 U_\nu= U_\alpha U_\beta U_\gamma U_\delta U_\epsilon\,,
 \label{eq:Nu_mixing_1A_nonD}
\end{equation}
with the respective pieces reported in Eq.~\eqref{eq:Nu_Umatrices_1A_nonD}. 
Note that now, there is no need to correct the mass ordering, since the $\delta_0^2$-term 
lowers the second eigenvalue down such that the mass square difference between the first and the 
second squared eigenvalues can now be as small as the measured value of $\Delta m_\odot^2$. The neutrino mass eigenvalues 
can again be determined as the absolute values of the diagonal entries of the resulting mass matrix. 
They turn out to be
\be
\left\{
\begin{array}{lcl}
 |m_1| &=& m_{\nu 0} \lambda^2 \; \frac{R_0^2}{2 \sqrt{R_0^4+2 R_0^2 \cos (2 \alpha_0 )+1}}+\mathcal{O}(\lambda^4)\,,\nonumber\\
 |m_2| &=& m_{\nu 0} \; \left(\delta_0^2+2 \lambda^2\right)+\mathcal{O}(\lambda^4)\,,\nonumber\\
 |m_3| &=& m_{\nu 0} \; (1+ 2 R_0^2 \lambda^2 )+\mathcal{O}(\lambda^4)\,.
 \label{eq:Nu_eigenvals_1A_nonD}
\end{array}
\right.
\ee
The mass square differences are, at lowest order,
\be
\left\{
\begin{array}{lcl}
 \Delta m_\odot^2 &\simeq& m_{\nu 0}^2 \delta_0^2 \; (\delta_0^2 + 4\lambda^2)+\mathcal{O}(\lambda^4)\,,\nonumber\\
 \Delta m_A^2 &\simeq& m_{\nu 0}^2 \; ( 1+ 4 R_0^2 \lambda ^2)+\mathcal{O}(\lambda^4)\,.
 \label{eq:Nu_Deltas_1A_nonD}
\end{array}
\right.
\ee
Using the numbers from Eq.~\eqref{eq:CL_ratios}, as well ad $\delta_0^2\simeq 0.18$, we now predict 
the ratio $\frac{\Delta m_\odot^2}{\Delta m_A^2}\simeq 0.034$, which is a quasi perfect match to the 
data. In fact, the trick is that the ratio between the mass square differences is just given 
by $\delta_0^4$, which precisely justifies the choice made for $\delta_0^2$. The PMNS-matrix turns out to be
\begin{equation}
 U_{\rm PMNS}^{\rm (1A), I}=\begin{pmatrix}
 1+\left(R_0^2+\frac{1}{\delta }\right) \lambda^2 & -\frac{i (\delta_0 -1) \lambda }{\delta_0 } & 0 \\
 \frac{(\delta_0 -1) \lambda }{\delta_0 } & i(1+\frac{\lambda^2}{\delta_0 }) 
 & u^{\rm (1A), I}_{23} \lambda^2\\
 0 & \frac{i R_0 e^{-i \alpha_0} \left(-1+\delta_0+e^{2i \alpha_0} \delta_0^3+\delta_0^4\right)}{\delta_0^4 -1} \lambda^2 
 & i \left(1+R_0^2 \lambda^2\right)
 \end{pmatrix}+ \mathcal{O}(\lambda^3)\,,
 \label{eq:PMNS_1A_nonD}
\end{equation}
where $u^{\rm (1A), I}_{23}=R_0 \left(-\frac{i (\delta^2_0+\delta_0-1) \cos \alpha_0 }{\delta^2_0 -1}+\sin \alpha_0- 
\frac{\delta_0 \sin \alpha_0}{\delta_0^2+1}\right)$. 
Although this matrix is still not really perfect, the value of the angle $\theta_{13}$ is 
much better than before, since $\theta_{13}\simeq \mathcal{O}(\lambda^3)$. 
Hence, we can be optimistic to find quasi-perfect numerical models when perturbing the non-democratic mass matrices.

Finally, for the non-democratic Scenario~B, we have a neutrino mixing matrix given by
\begin{equation}
 U_\nu= U'_\alpha U'_\beta U'_\gamma U'_\delta U'_\epsilon,
 \label{eq:Nu_mixing_1B_nonD}
\end{equation}
with the respective pieces given in Eq.~\eqref{eq:Nu_Umatrices_1B_nonD}. 
Also in this case, there is no need to correct for inverted ordering. The eigenvalues read
\be
\left\{
\begin{array}{lcl}
 |m_1| &=& m_{\nu 0} \lambda^2 \; \frac{\sqrt{R_0^4+2 R_0^2 \cos (2 \alpha_0 )+1}}{2 R_0^2}+\mathcal{O}(\lambda^4)\,,\nonumber\\ 
 |m_2| &=& m_{\nu 0} \; \frac{\left(R_0^4 \left(2 \delta_0^2+\lambda^2\right)-2 R_0^2 \lambda^2 \cos (2 \alpha_0 )
+\lambda^2\right) }{R_0^2 \sqrt{R_0^4-2 R_0^2 \cos (2 \alpha_0 )+1}}+\mathcal{O}(\lambda^4)\,,\nonumber\\
 |m_3| &=& m_{\nu 0} \; (1+ 2 R_0^2 \lambda^2)+\mathcal{O}(\lambda^4)\,,
 \label{eq:Nu_eigenvals_1B_nonD}
\end{array}
\right.
\ee
and the mass square differences are, at lowest order,
\be
\left\{
\begin{array}{lcl}
 \Delta m_\odot^2 &\simeq& 4 m_{\nu 0}^2 \delta_0^2 \; \left( \lambda^2+\frac{\delta_0^2 R_0^4}{R_0^4 - 2 R_0^2 
\cos (2 \alpha_0 )+1} \right)+\mathcal{O}(\lambda^4)\,,\nonumber\\
 \Delta m_A^2 &\simeq& m_{\nu 0}^2 \; (1+ 4 R_0^2 \lambda^2 )+\mathcal{O}(\lambda^4)\,.
 \label{eq:Nu_Deltas_1B_nonD}
\end{array}
\right.
\ee
Again taking the most suitable value of $\alpha_0=\frac{\pi}{2}$, the ratio of mass square differences, 
$\frac{\Delta m_\odot^2}{\Delta m_A^2}$, is now 0.046, which is still too large but much closer to the 
actual value than before. The PMNS matrix is given by
\begin{equation}
 U_{\rm PMNS}^{\rm (1B), I}=\begin{pmatrix}
 1+\left(R_0^2+\frac{1}{2 \delta_0 }-\frac{e^{-2 i \alpha_0 }}{2 \delta_0 R_0^2}\right) \lambda^2 
& -\frac{i \left((2 \delta_0 -1) R_0^2+e^{2 i\alpha_0}\right)}{2 R_0^2 \delta_0}\lambda & 0 \\
 \left(1+\frac{e^{-2 i \alpha_0 }-R_0^2}{2 R_0^2\delta_0 }\right) \lambda  
&  i \left(1+\frac{\left(R_0^2-e^{2 i \alpha_0 }\right)}{2 R_0^2 \delta_0 }\right) \lambda^2 & u^{\rm (1B), I}_{23} \lambda^2\\
 0 & u^{\rm (1B), I}_{32} \lambda^2 & i \left(1+R_0^2 \lambda^2\right)
 \end{pmatrix}+ \mathcal{O}(\lambda^3)\,,
 \label{eq:PMNS_1B_nonD}
\end{equation}
where
\begin{eqnarray}
 u^{\rm (1B), I}_{23} &=& R_0 \left(-i e^{-i \alpha_0}+\frac{n_{23}}
{\left(4 \delta_0^4-1\right) R_0^4+2 \cos(2 \alpha_0) R_0^2-1}\right),\nonumber\\
 n_{23} &=& \delta_0 \left(2 e^{-2 i \alpha_0} \delta_0^2 R_0^2
+R_0^2-e^{2 i \alpha_0}\right) \left(\left(R_0^2+1\right) \sin \alpha_0 -i \left(R_0^2-1\right) \cos \alpha_0 \right),\nonumber\\
 u_{32}^{\rm (1B), I} &=& \frac{i e^{-3 i \alpha_0} R_0 n_{32}}{\left(4 \delta_0^4-1\right) R_0^4+2 \cos (2 \alpha_0) R_0^2-1},
\nonumber\\
 n_{32} &=& -2 e^{6 i\alpha_0} R_0^2 \delta_0^3-R_0^2 (\delta_0 -1)
 +e^{4 i \alpha_0} R_0^2 \left(2 R_0^2 \delta_0^3-\delta_0 +1\right)+\nonumber\\
 && +e^{2 i \alpha_0} \left(\left(4 \delta_0^4+\delta_0-1\right) R_0^4 
 +\delta_0 -1\right).
 \label{eq:u_non-dem}
\end{eqnarray}
Also in this case, the value of the angle $\theta_{13}$ is 
much better than in the democratic case, since 
$\theta_{13}\simeq \mathcal{O}(\lambda^3$). 
However, to find quasi-perfect models, we will have to turn to a numerical study, which 
is presented in the next section.

%%%%%%%%%%%%%%%%%%%%%%%%%%%%%%%%%%%%%%%%%%%%%%%%%%%%%%%%%%%%%%%%%%%%%%
\subsection{\label{sec:numerical} Numerical analysis}
%%%%%%%%%%%%%%%%%%%%%%%%%%%%%%%%%%%%%%%%%%%%%%%%%%%%%%%%%%%%%%%%%%%%%%

\begin{table}[t]
 \centering
\begin{tabular}{|c||c|}\hline
1AI & Matrix \\ \hline \hline
$M_e$ &  $M_{e0} \left(
\begin{array}{lll}
 0.81 B_2 \lambda^3 & 1.44 B_2 \lambda^2 & 0.29 R \lambda^2 \\
 2.00 B_2 \lambda^2 & 1.13 \lambda  & 2.50 R \lambda  \\
 3.71 R \lambda  & 0 & 0.35
\end{array}
\right)$\\ \hline
$m_D$ & $m_{D0} \left(
\begin{array}{lll}
 0.75 B_4 \lambda^4 & 0.15 B_2 \lambda^3 & 1.42 B_2 R \lambda^3 \\
 0.51 \lambda  & 0.13 & 0 \\
 3.32 R \lambda  & 0 & 2.93
\end{array}
\right)$\\ \hline
$M_R$ & $M_0 \left(
\begin{array}{lll}
 0.38 B_6 \lambda^6 & 0.31 B_2 \lambda^3 & 1.26 B_2 R \lambda^3 \\
 0.31 B_2 \lambda^3 & 4.18 & 0 \\
 1.26 B_2 R \lambda^3 & 0 & 4.81
\end{array}
\right)$\\ \hline
\end{tabular}
 \caption{\label{tab:1AI} 
Mass matrices of model 1AI. The prefactors parametrize the mass scales imposed on the concrete model at a certain energy scale.}
\end{table}

\begin{table}[t]
 \centering
\begin{tabular}{|c||c|}\hline
1BI & Matrix \\ \hline \hline
$M_e$ &  $M_{e0} \left(
\begin{array}{lll}
 0.91 B_2 \lambda^3 & 2.26 B_2 \lambda^2 & 4.33 R \lambda^2 \\
 3.80 B_2 \lambda^2 & 2.51 \lambda  & 3.63 R \lambda  \\
 0.79 R \lambda  & 0 & 0.15
\end{array}
\right)$\\ \hline
$m_D$ & $m_{D0} \left(
\begin{array}{lll}
 3.20 B_4 \lambda^5 & 0.15 B_4 \lambda^4 & 3.73 B_2 R \lambda^4 \\
 2.27 B_2 \lambda^2 & 0.030 \lambda  & 1.23 R \lambda  \\
 1.69 R \lambda  & 0 & 0.66
\end{array}
\right)$\\ \hline
$M_R$ & $M_0 \left(
\begin{array}{lll}
 0.33 B_8 \lambda^8 & 2.68 B_4 \lambda^5 & 1.63 B_2 R \lambda^4 \\
 2.68 B_4 \lambda^5 & 2.10 B_2 \lambda^2 & 0.83 R \lambda  \\
 1.63 B_2 R \lambda^4 & 0.83 R \lambda  & 0.49
\end{array}
\right)$\\ \hline
\end{tabular}
 \caption{\label{tab:1BI} 
Mass matrices of model 1BI. The prefactors parametrize the mass scales imposed on the concrete model at a certain energy scale.}
\end{table}

Finally, we want to present a few quasi-perfect models and their predictions. Usually, and in particular in the context of 
FN inspired models, so-called scatter plots are presented in the literature. These plots are supposed to indicate that the 
model is consistent with data in a considerable region of the parameter space. However, we do not consider this approach as 
too useful, since first such results always depend on the statistical measure used in the generation of the random numbers 
involved, and second because there is essentially no method to determine if indeed a large part of the parameter space is 
investigated or rather only a tiny patch, due to the gigantic complexity of higher-dimensional non-Cartesian spaces. 
Furthermore, there is no way to decide whether a model can be considered as ``good'' or ``bad'', just because a certain 
fraction of random parameter choices leads to compatibility with data.

We will rather take on a contrary approach and try to find four fully working examples by perturbing the mass matrices from 
Eqs.~\eqref{eq:charged_matrices_12}, \eqref{eq:MR_matrices_AB}, \eqref{eq:mD_matrices_1AB}, and~\eqref{eq:mD_matrices_2AB}, 
where the Dirac Yukawa coupling matrices $Y_D$ are taken to be non-democratic, according to Sec.~\ref{sec:non-democratic}. 
If these examples are in agreement with data, we consider the FN approach as being predictive in the sense that small 
departures from the analytical forms of the mass matrices are perfectly enough to be in full agreement with low-energy 
neutrino data.

Based on the analytical results from Sec.~\ref{sec:democratic}, we choose the parameters $\lambda$, $R_0$, and $\alpha_0$ 
in the following way:
\begin{itemize}
 \item Assignment 1: $\lambda=0.06$, $R_0=1.18$, $\alpha_0=0.67$,

 \item Assignment 2: $\lambda=0.10$, $R_0=0.80$, $\alpha_0=0.60$.

\end{itemize}
The next step is to generate a number of perturbed charged lepton matrices according to Eq.~\eqref{eq:charged_matrices_12}, and 
check which of them yield a quasi perfect prediction ($\pm 1\%$) for the charged lepton mass ratios $m_e/m_\mu$ and 
$m_\mu/m_\tau$. We furthermore generate right-handed neutrino mass matrices according to Eq.~\eqref{eq:MR_matrices_AB}, whose 
smallest eigenvalue should always be at least six orders of magnitude smaller than the one of the second to lightest mass 
eigenstate. We then use these matrices together with the Dirac mass matrices from Eqs.~\eqref{eq:mD_matrices_1AB} 
and~\eqref{eq:mD_matrices_2AB} to generate light neutrino mass matrices according to the seesaw formula, 
$m_\nu=-m_D^T M_R^{-1} m_D$, which yield a quasi-perfect prediction ($\pm 1\%$) for the ratio of neutrino mass square differences, 
$\Delta m_\odot^2/\Delta m_A^2$. Finally we combine the lists of matrices and calculate the mixing parameters and phases using 
MPT~\cite{Antusch:2005gp,MPT}. All models that lead to predictions of the mixing angles that are consistent with the current 
$3\sigma$ ranges~\cite{Schwetz:2011qt} will survive the test.

The corresponding mass matrices of the quasi-perfect numerical models for all four combinations of Scenarios~A and~B, as well as of 
Assignments~1 and 2 can be found in Tabs.~\ref{tab:1AI} to~\ref{tab:2BI}. Indeed, all mass matrices have just the structure 
that we have desired: Only slight perturbations of the democratic form, with the exception of the Dirac mass matrices that 
have smaller 12 and 22 entries. Accordingly, as we had expected, the FN models presented here are predictive up to the exact 
values of the mixing angles and $CP$ phases. In particular, all charged lepton and light neutrino mass matrices lead to the 
correct ratios of masses or mass square differences, respectively.

\begin{table}[t]
 \centering
\begin{tabular}{|c||c|}\hline
2AI & Matrix \\ \hline \hline
$M_e$ &  $M_{e0} \left(
\begin{array}{lll}
 4.96 B_4 \lambda^4 & 3.70 B_2 \lambda^3 & 3.63 B_2 R \lambda^3 \\
 4.75 B_2 \lambda^3 & 1.51 B_2 \lambda^2 & 3.42 R \lambda^2 \\
 3.17 R \lambda^2 & 0.70 R \lambda  & 0.28 \lambda 
\end{array}
\right)$\\ \hline
$m_D$ & $m_{D0} \left(
\begin{array}{lll}
 3.79 B_4 \lambda^5 & 0.0057 B_4 \lambda^4 & 4.60 B_2 R \lambda^4 \\
 0.68 B_2 \lambda^2 & 0.097 \lambda  & 1.20 R \lambda  \\
 1.12 R \lambda^2 & 3.30 R \lambda  & 4.85 \lambda
\end{array}
\right)$\\ \hline
$M_R$ & $M_0 \left(
\begin{array}{lll}
 3.02 B_6 \lambda^6 & 1.28 B_2 \lambda^3 & 0.93 B_2 R \lambda^3 \\
 1.28 B_2 \lambda^3 & 4.32 & 0 \\
 0.93 B_2 R \lambda^3 & 0 & 3.08
\end{array}
\right)$\\ \hline
\end{tabular}
 \caption{\label{tab:2AI} 
Mass matrices of model 2AI. The prefactors parametrize the mass scales imposed on the concrete model at a certain energy scale.}
\end{table}

\begin{table}[t]
 \centering
\begin{tabular}{|c||c|}\hline
2BI & Matrix \\ \hline \hline
$M_e$ &  $M_{e0} \left(
\begin{array}{lll}
 2.71 B_4 \lambda^4 & 4.17 B_2 \lambda^3 & 3.74 B_2 R \lambda^3 \\
 1.39 B_2 \lambda^3 & 0.19 B_2 \lambda^2 & 3.19 R \lambda^2 \\
 0.91 R \lambda^2 & 0.38 R \lambda  & 0.40 \lambda 
\end{array}
\right)$\\ \hline
$m_D$ & $m_{D0} \left(
\begin{array}{lll}
 1.57 B_6 \lambda^6 & 0.078 B_4 \lambda^5 & 1.79 B_4 R \lambda^5 \\
 4.87 B_2 \lambda^3 & 0.16 B_2 \lambda^2 & 4.85 R \lambda^2 \\
 3.47 R \lambda^2 & 1.82 R \lambda  & 1.25 \lambda 
\end{array}
\right)$\\ \hline
$M_R$ & $M_0 \left(
\begin{array}{lll}
 0.61 B_8 \lambda^8 & 1.12 B_4 \lambda^5 & 0.53 B_2 R \lambda^4 \\
 1.12 B_4 \lambda^5 & 3.85 B_2 \lambda^2 & 4.33 R \lambda \\
 0.53 B_2 R \lambda^4 & 4.33 R \lambda  & 3.21
\end{array}
\right)$\\ \hline
\end{tabular}
 \caption{\label{tab:2BI} 
Mass matrices of model 2BI. The prefactors parametrize the mass scales imposed on the concrete model at a certain energy scale.}
\end{table}

\begin{figure}[t]
\centering
\begin{tabular}{cc}
\includegraphics[width=6.5cm,height=6cm]{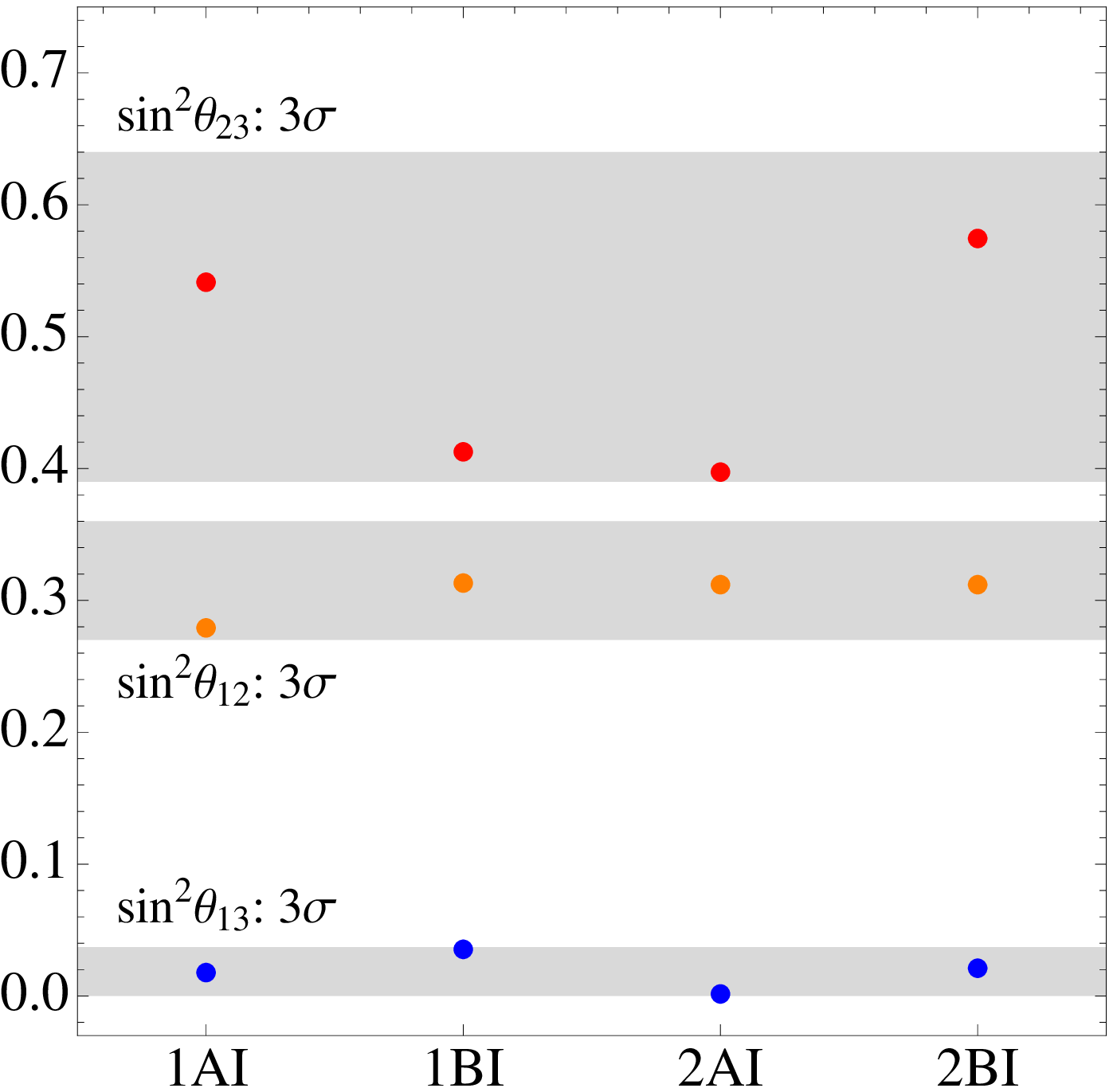}& 
\includegraphics[width=7.5cm,height=6cm]{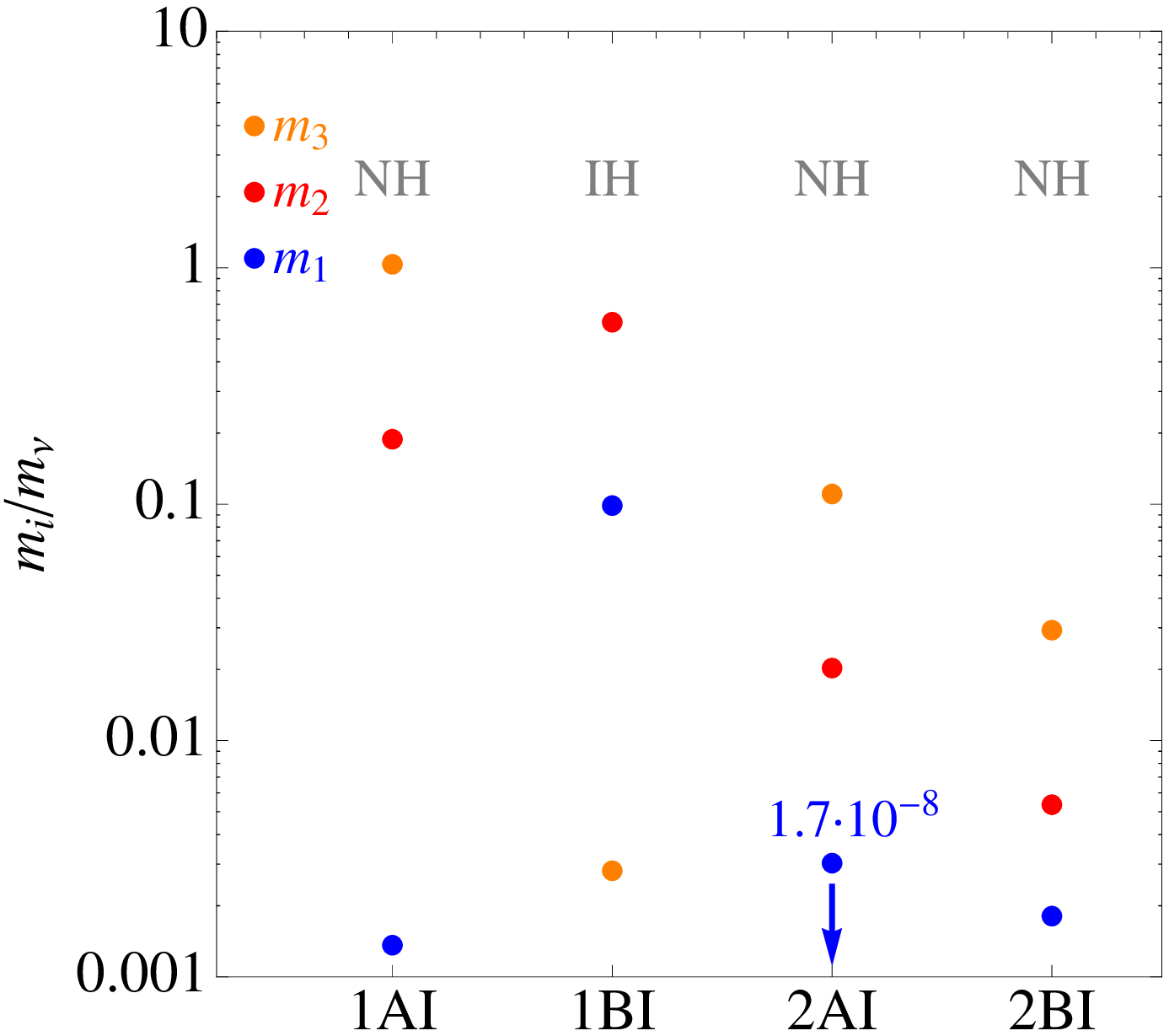}
\end{tabular}
\caption{\label{fig:angles_masses} The predictions of the numerical models for the leptonic mixing angles as for the masses, 
the latter in units of the light neutrino mass scale $m_\nu$.}
\end{figure}

The numerical predictions are summarized in Tab.~\ref{tab:predictions}, and the mixing angles and masses are also displayed 
in Fig.~\ref{fig:angles_masses}. As can be seen, all mixing angle predictions are in agreement with the $3\sigma$ ranges of the 
oscillation 
parameters~\cite{Schwetz:2011qt}, because the models have been selected in that way. Furthermore, as explained in 
Sec.~\ref{sec:RGE}, these predictions are effectively scale invariant. However, a non-trivial prediction is the actual values 
of the masses (in terms of the absolute neutrino mass scale $m_\nu$, which is determined by the other scales involved), and in 
particular the dominant prediction of normal mass ordering (or, due to the hierarchical structure of FN matrices, normal 
hierarchy), which we had already anticipated in Sec.~\ref{sec:non-democratic}.

\begin{table}[t]
 \centering
\begin{tabular}{|l||c|c|c||c|c|c||c|c|c||c|}\hline
Mod. & $s^2_{12}$ & $s^2_{13}$ & $s^2_{23}$ & $\delta$ & $\alpha$ & $\beta$ & $m_1/m_\nu$ & $m_2/m_\nu$ & $m_3/m_\nu$ & Hier. 
\\ \hline \hline
1AI & 0.28 & 0.018 & 0.54 & 4.21 & 0.58 & 1.25 & 0.0014 & 0.19 & 1.03 & NH \\ \hline
1BI & 0.31 & 0.035 & 0.41 & 4.98 & 2.35 & 2.89 & 0.099 & 0.59 & 0.0028 & IH  \\ \hline
2AI & 0.31 & 0.0015 & 0.40 & 3.88 & 2.27 & 0.54 & $1.7\cdot 10^{-8}$ & 0.020 & 0.11 & NH  \\ \hline
2BI & 0.27 & 0.021 & 0.57 & 3.83 & 1.96 & 2.71 & 0.0018 & 0.0054 & 0.029 & NH  \\ \hline
\end{tabular}
 \caption{\label{tab:predictions} 
Predictions of the four numerical models. ``Mod.'' stands for ``model'', $s^2_{ij}\equiv \sin^2 \theta_{ij}$, and ``Hier.'' 
stands for ``hierarchy''.}
\end{table}

%%%%%%%%%%%%%%%%%%%%%%%%%%%%%%%%%%%%%%%%%%%%%%%%%%%%%%%%%%%%%%%%%%%%%%
\section{\label{sec:conc} Conclusions}
%%%%%%%%%%%%%%%%%%%%%%%%%%%%%%%%%%%%%%%%%%%%%%%%%%%%%%%%%%%%%%%%%%%%%%

In this paper, we have shown how to carefully derive models explaining the appearance of one 
right-handed (sterile) neutrino that has a mass at the keV scale, while at the same time predicting 
leptonic mass ranges and mixing parameters in full agreement with experiments. These models were 
based on the famous Froggatt-Nielsen mechanism, which is a well-known possibility to create strong 
hierarchies between fermion masses. 
To our knowledge, this is the third known type of models that can successfully
explain the existence of a keV sterile neutrino Dark Matter particle, the first two
being based on soft breaking of flavour symmetries~\cite{Shaposhnikov:2006nn, Lindner:2010wr} and on Randall-Sundrum
warping~\cite{Randall:1999ee}, respectively, where the former exploits symmetry breaking effects to
lift a massless state to the keV scale while the latter uses the exponential
suppression of UV-brane physics to strongly suppress the natural right-handed
neutrino mass scale. 
We have instead made use of the possibility to assign FN charges to the different fermions in such a 
way as to suppress certain mass matrix elements, whose sizes are reduced by roughly one order of magnitude 
per unit FN charge. In this way, it is possible to suppress one right-handed neutrino mass strongly enough 
to be around the keV scale, while the other two can easily be considerably heavier (a mass of about GeV is 
the lower bound to be fulfilled, but they could easily be much heavier). One important bonus of this 
approach is that, due to the structure of the FN charge assignments, the seesaw mechanism will be guaranteed 
to work if it works with all FN charges set to zero, so that we can be sure that the existence of a keV particle 
will not lead to any problems from that side. Furthermore, we discuss the requirements and potential problems 
that could arise when applying the FN mechanism in certain frameworks. Interestingly, although FN charge 
assignments might seem relatively arbitrary at the first sight, it is easy to find situations where the 
different conditions are so restrictive as to render the different sectors incompatible. This discussion leads 
to a systematic reduction of the arbitrariness involved, and it turns out that an ideal framework for our purpose 
is an $SU(5)$ GUT, augmented by two FN flavon fields. Due to the relatively low seesaw scale involved, typical 
constraints from RGE running and the corresponding LFV effects can be avoided. Finally, we show analytically 
that democratic structures of the mass matrices are not enough to explain all mass ratios, but once we depart 
from this structure only slightly, the agreement with the data improves. In order to find fully working models, which are 
also in agreement with the experimental constraints on the leptonic mixing angles, we finally perform a numerical 
analysis that perfectly justifies our argumentations given before.

%%%%%%%%%%%%%%%%%%%%%%%%%%%%%%%%%%%%%%%%%%%%%%%%%%%%%%%%%%%%%%%%%%%%%%
\section*{\label{sec:Ack} Acknowledgements}
%%%%%%%%%%%%%%%%%%%%%%%%%%%%%%%%%%%%%%%%%%%%%%%%%%%%%%%%%%%%%%%%%%%%%%

We would like to thank J.~Barry, J.~Bergstr\"om, C.~Luhn, K.~L.~McDonald, and H.~Zhang for useful discussions, and we are especially 
grateful to Sheldon, Leonard, Howard, and Rajesh for keeping our motivation always at 
the top level. The work of AM is supported by the Royal Institute 
of Technology (KTH), under project no.\ SII-56510. 
VN acknowledges Research Grants funded jointly by Ministero
dell'Istruzione, dell'Universit\`a e della Ricerca (MIUR), by
Universit\`a di Torino, and by Istituto Nazionale di Fisica Nucleare
within the {\sl Astroparticle Physics Project} (MIUR contract number:
PRIN 2008NR3EBK; INFN grant code: FA51).

% =============================================================================
\section*{\label{app:diagonal}Appendix A: Stepwise diagonalization}
% =============================================================================

\renewcommand{\theequation}{A-\arabic{equation}}
% redefine the command that creates the equation no.
\setcounter{equation}{0}  % reset counter
We report in this appendix the explicit expressions of the approximately unitary 
mixing matrices that diagonalize the charged leptons mass matrix and the 
neutrino mass matrices for Assignment~1, Scenarios~A and~B, with both 
democratic and non-democratic Dirac Yukawa matrices.

% =============================================================================
\subsection*{Charged leptons}
\label{app:charged}
% =============================================================================

Here we report the mixing matrices that diagonalize the charged leptons mass 
matrix, as reported in Eq.~\eqref{eq:CL_mixing}:
\begin{eqnarray}
 && U_A=\begin{pmatrix}
 1 & 0 & a_1 \lambda  \\
 0 & 1 & 0 \\
 -a_1^* \lambda  & 0 & 1
 \end{pmatrix},\ \
 U_B=\begin{pmatrix}
 1 & 0 & 0 \\
 0 & 1 & b_1 \lambda^2 \\
 0 & -b_1^* \lambda^2 & 1
 \end{pmatrix},\ \
 U_C=\begin{pmatrix}
 1 & 0 & c_1 \lambda^3 \\
 0 & 1 & 0 \\
 -c_1^* \lambda^3 & 0 & 1
 \end{pmatrix},\nonumber \\
 && U_D=\begin{pmatrix}
 1 & \lambda  & 0 \\
 -\lambda  & 1 & 0 \\
 0 & 0 & 1
 \end{pmatrix},\ \
 U_E=\begin{pmatrix}
 1 & 0 & 0 \\
 0 & 1 & e_1 \lambda^4 \\
 0 & -e_1^* \lambda^4 & 1
 \end{pmatrix},
 U_F=\begin{pmatrix}
 1 & 0 & f_1 \lambda^5 \\
 0 & 1 & 0 \\
 -f_1^* \lambda^5 & 0 & 1
 \end{pmatrix},\nonumber\\
 && U_G=\begin{pmatrix}
 1 & g_1 \lambda^3 & 0 \\
 -g_1^* \lambda^3 & 1 & 0 \\
 0 & 0 & 1
 \end{pmatrix}, 
 \label{eq:CL_Umatrices}
\end{eqnarray}
with
\begin{eqnarray}
 a_1=b_1^*=c_1&=& R_0 e^{-i \alpha_0 },\nonumber\\
 e_1&=& R_0 e^{i \alpha_0 } \left(R_0^2 e^{-2 i \alpha_0 }-2 R_0^2+e^{-2 i \alpha_0 }+2\right),\nonumber\\
 f_1&=& -R_0^3 e^{-i \alpha_0 },\nonumber\\
 g_1&=& -R_0^2 \left(R_0^2+e^{-2 i \alpha_0 }\right).
 \label{eq:CL_entries}
\end{eqnarray}

% =============================================================================
\subsection*{Light neutrinos, Assignment~1, Scenario~A, democratic}
\label{app:light1A}
% =============================================================================

Here we report the mixing matrices that diagonalize the light neutrino mass 
matrix, see Eq.~\eqref{eq:Nu_mixing_1A}, for Assignment~1, Scenario~A, and democratic Yukawa matrices:
\begin{eqnarray}
 && U_\alpha=\begin{pmatrix}
 1 & 0 & \alpha_1 \lambda  \\
 0 & 1 & 0 \\
 -\alpha_1^* \lambda  & 0 & 1
 \end{pmatrix},\ \
 U_\beta=\begin{pmatrix}
 1 & \lambda & 0 \\
 -\lambda & 1 & 0 \\
 0 & 0 & 1
 \end{pmatrix},\ \
 U_\gamma=\begin{pmatrix}
 1 & 0 & 0 \\
 0 & 1 & \gamma_1 \lambda^2 \\
 0 & -\gamma^*_1 \lambda^2 & 1
 \end{pmatrix},\nonumber \\
 && U_\delta=\begin{pmatrix}
 1 & 0 & \delta_1\lambda^3 \\
 0 & 1 & 0 \\
 -\delta^*_1\lambda^3 & 0 & 1
 \end{pmatrix},\ \
 U_\epsilon=\begin{pmatrix}
 1 & \epsilon_1 \lambda^3 & 0 \\
 -\epsilon_1^* \lambda^3 & 1 & 0 \\
 0 & 0 & 1
 \end{pmatrix},
 U_\zeta=\begin{pmatrix}
  1 & 0 & 0 \\
 0 & \frac{\zeta_1}{\sqrt{R_0^2+ \zeta_1^2}} & -\frac{\zeta_1}{\sqrt{R_0^2+ \zeta_1^2}}\\
 0 & \frac{1}{\sqrt{R_0^2+ \zeta_1^2}} & \frac{1}{\sqrt{R_0^2+ \zeta_1^2}}
 \end{pmatrix}\,,\nonumber\\
 && U_\eta=\begin{pmatrix}
 0 & 0 & 1 \\
 0 & 1 & 0 \\
 1 & 0 & 0
 \end{pmatrix},
 \label{eq:Nu_Umatrices_1A}
\end{eqnarray}
with
\begin{eqnarray}
 \alpha_1 &=& R_0 e^{-i \alpha_0 }\,,\nonumber\\
 \gamma_1 &=& \frac{i}{2} R_0 \sin \alpha_0\,,\nonumber\\
 \delta_1 &=& -\frac{e^{-i \alpha_0} R_0^3}{2(1+e^{-2 i \alpha_0} R_0^2)}\,,\nonumber\\
 \epsilon_1&=& -\frac{e^{-2 i \alpha_0} R_0^2}{2(1+e^{-2 i \alpha_0} R_0^2)}\,,\nonumber\\
 \zeta_1&=& \left(1-R_0^2+\sqrt{(-1+R_0^2)^2 +R_0^2 \cos^2 \alpha_0}\right) \sec \alpha_0\,.
 \label{eq:Nu_entries_1A}
\end{eqnarray}

% =============================================================================
\subsection*{Light neutrinos, Assignment~1, Scenario~B, democratic}
\label{app:light1B}
% =============================================================================

Here we report the mixing matrices that diagonalize the light neutrino mass 
matrix, see Eq.~\eqref{eq:Nu_mixing_1B}, for Assignment~1, Scenario~B, and democratic Yukawa matrices:
\begin{eqnarray}
\label{eq:Nu_Umatrices_1B} 
 && U'_\alpha=\begin{pmatrix}
 1 & 0 & \alpha'_1 \lambda  \\
 0 & 1 & 0 \\
 -(\alpha'_1)^* \lambda  & 0 & 1
 \end{pmatrix},\ \
 U'_\beta=\begin{pmatrix}
 1 & \beta'_1 \lambda & 0 \\
 -(\beta'_1)^* \lambda & 1 & 0 \\
 0 & 0 & 1
 \end{pmatrix},\ \
 U'_\gamma=\begin{pmatrix}
 1 & 0 & 0 \\
 0 & i & i \gamma'_1 \lambda^2\\
 0 & -i (\gamma'_1)^* \lambda^2 & i
 \end{pmatrix},\nonumber \\
 && U'_\delta=\begin{pmatrix}
 1 & 0 & \delta'_1 \lambda^3 \\
 0 & 1 & 0 \\
 -(\delta'_1)^* \lambda^3 & 0 & 1
 \end{pmatrix}\,,\ \
 U'_\epsilon=\begin{pmatrix}
 1 & \epsilon'_1 \lambda^3 & 0 \\
 -(\epsilon'_1)^* \lambda^3 & 1 & 0 \\
 0 & 0 & 1
 \end{pmatrix}\,,\ \
 U'_\zeta=\begin{pmatrix}
 0 & 0 & 1 \\
 0 & 1 & 0 \\
 1 & 0 & 0
 \end{pmatrix},
\end{eqnarray}
with
\begin{eqnarray}
 \alpha'_1 &=& R_0 e^{-i \alpha_0 },\nonumber\\
 \beta'_1&=& \frac{ R_0^2-e^{2 i \alpha_0 }}{2 R_0^2},\nonumber\\
 {\rm Re}(\gamma'_1)&=& -\frac{R_0 \cos \alpha_0 [1 + 2 R_0^2 + 3 R_0^4 - 
   6 R_0^2 \cos(2 \alpha_0)]}{-1 + 3 R_0^4 + 2 R_0^2 \cos(2 \alpha_0)}\,.\nonumber\\
 {\rm Im}(\gamma'_1)&=&\frac{R_0 \sin \alpha_0 [-1 - 2 R_0^2 + R_0^4 -2 R_0^2 
\cos(2 \alpha_0)]}{-1 + 3 R_0^4 + 2 R_0^2 \cos(2 \alpha_0)}\,,\nonumber\\
 \delta'_1 &=& \frac{i e^{i \alpha_0} (e^{2 i \alpha_0}+R_0^2)}{2 R_0}\,,\nonumber\\
 \epsilon'_1 &=& \frac{i (-e^{2 i \alpha_0} + R_0^2)[-1+R_0^4+2 i R_0^2 \sin(2\alpha_0)]}{8 R_0^6}\,.
\label{eq:Nu_entries_1B}
\end{eqnarray}

% =============================================================================
\subsection*{Light neutrino, Assignment~1, Scenario~A, non-democratic}
\label{app:light1AnonD}
% =============================================================================

Here we report the mixing matrices that diagonalize the light neutrino mass 
matrix, see Eq.~\eqref{eq:Nu_mixing_1A_nonD}, for Assignment~1, Scenario~A, and 
non-democratic Yukawa matrices:
\begin{eqnarray}
 && U_\alpha=\begin{pmatrix}
 1 & 0 & \alpha_1 \lambda  \\
 0 & 1 & 0 \\
 -\alpha_1^* \lambda  & 0 & 1
 \end{pmatrix},\ \
 U_\beta=\begin{pmatrix}
 1 & 0 & 0 \\
 0 & i & i\beta_1 \lambda^2 \\
 0 & -i\beta_1^* \lambda^2 & i
 \end{pmatrix},\ \
 U_\gamma=\begin{pmatrix}
 1 & \frac{i \lambda }{\delta_0 } & 0 \\
 \frac{i \lambda }{\delta_0 } & 1 & 0 \\
 0 & 0 & 1
 \end{pmatrix},\nonumber \\
 && U_\delta=\begin{pmatrix}
 1 & 0 & \delta_1 \lambda^3 \\
 0 & 1 & 0  \\
 -\delta_1^* \lambda^3 & 0 & 1
 \end{pmatrix}\,,\ \ 
 U_\epsilon=\begin{pmatrix}
 1 & \epsilon_1 \lambda^3 & 0 \\
 -\epsilon_1^* \lambda^3 & 1 & 0  \\
 0 & 0 & 1
 \end{pmatrix},
 \label{eq:Nu_Umatrices_1A_nonD}
\end{eqnarray}
with
\begin{eqnarray}
 \alpha_1 &=& R_0 e^{-i \alpha_0 }\,,\nonumber\\
 { \rm Re}(\beta_1)&=& -\frac{R_0 \delta_0 \cos \alpha_0 }{-1+\delta^2}\,,\nonumber\\
 {\rm Im}(\beta_1)&=& \frac{R_0 \delta_0 \sin \alpha_0 }{1+\delta^2}\,,\nonumber\\
 \delta_1&=& -\frac{i e^{-i\alpha_0} R_0 \left(2 e^{4 i \alpha_0}+2 R_0^2 \delta_0^2+
e^{2 i \alpha_0} [2\delta_0^2+R_0^2(1+\delta_0^4)] \right)}{2 (e^{2 i \alpha_0}+R_0^2) (-1+\delta_0^4)}\,,\nonumber\\
 \epsilon_1&=&\frac{i e^{2 i \alpha_0} R_0^2}{2 \delta_0^3 (1+e^{2 i \alpha_0} R_0^2)}\,.
 \label{eq:Nu_entries_1A_nonD}
\end{eqnarray}

% =============================================================================
\subsection*{Light neutrinos, Assignment~1, Scenario~B, non-democratic}
\label{app:light1BnonD}
% =============================================================================

Here we report the mixing matrices that diagonalize the light neutrino mass 
matrix, see Eq.~\eqref{eq:Nu_mixing_1B_nonD}, for Assignment~1, Scenario~B, and 
non-democratic Yukawa matrices:
\begin{eqnarray}
 && U'_\alpha=\begin{pmatrix}
 1 & 0 & \alpha'_1 \lambda  \\
 0 & 1 & 0 \\
 -(\alpha'_1)^* \lambda  & 0 & 1
 \end{pmatrix},\ \
 U'_\beta=\begin{pmatrix}
 1 & \beta'_1 \lambda & 0 \\
 -(\beta'_1)^* \lambda & 1 & 0 \\
 0 & 0 & 1
 \end{pmatrix},\ \
 U'_\gamma=\begin{pmatrix}
 1 & 0 & 0 \\
 0 & i & i \gamma'_1 \lambda^2\\
 0 & -i (\gamma'_1)^* \lambda^2 & i
 \end{pmatrix},\nonumber\\ 
 && U'_\delta=\begin{pmatrix}
 1 & 0 & \delta'_1 \lambda^3 \\
 0 & 1 & 0 \\
 -(\delta'_1)^* \lambda^3 & 0 & 1
 \end{pmatrix},\ \
U'_\epsilon=\begin{pmatrix}
 1 & \epsilon'_1 \lambda^3 & 0 \\
 -(\epsilon'_1)^* \lambda^3 & 1 & 0 \\
 0 & 0 & 1
 \end{pmatrix},
 \label{eq:Nu_Umatrices_1B_nonD}
\end{eqnarray}
with
\begin{eqnarray}
 \alpha'_1 &=& R_0 e^{-i \alpha_0}\,,\nonumber\\
 \beta'_1&=& \frac{1}{2 \delta_0}-\frac{e^{2 i \alpha_0}}{2 \delta_0 R_0^2 }\,,\nonumber\\
 {\rm Re}(\gamma'_1)&=& \frac{R_0 \delta_0\left(-[1-R_0^2+R_0^4(1+2 \delta_0^2)]\cos \alpha_0 +
R_0^2 (1+2\delta_0^2)\cos(3 \alpha_0)\right)}{-1+R^4_0(-1+4\delta_0^4)+2R_0^2 \cos(2 \alpha_0)}\,,\nonumber\\
 {\rm Im}(\gamma'_1)&=& \frac{R_0 \delta_0\left(-[1-R_0^2+R_0^4(1+2 \delta_0^2)]\sin \alpha_0 +
R_0^2 (1-2\delta_0^2)\sin(3 \alpha_0)\right)}{-1+R^4_0(-1+4\delta_0^4)+2R_0^2 \cos(2 \alpha_0)}\,,\nonumber\\
 \delta'_1&=&\frac{i e^{3 i \alpha_0}(1+e^{-2 i \alpha_0}R_0^2)}{2 R_0}\,,\nonumber\\
 \epsilon'_1&=&\frac{i(R_0^6+ e^{2i\alpha_0}-R_0^4 e^{-2 i \alpha_0} -
R_0^2 e^{4i \alpha_0})}{8 R_0^6 \delta_0^3}\,.\nonumber\\
 \label{eq:Nu_entries_1B_nonD}
\end{eqnarray}

%=============================================================================
\bibliographystyle{./apsrev}
\bibliography{./keV_FN}
%=============================================================================

\end{document}